\documentclass[a4paper,onecolumn,draftcls,11pt]{IEEEtran}

\usepackage{amsmath}
\usepackage{amssymb}
\usepackage[noadjust]{cite}
\usepackage{paralist}
\usepackage{graphicx}
\usepackage{tikzexternal}
\author{David Gregoratti*,~\IEEEmembership{Senior Member,~IEEE,}
and Xavier~Mestre,~\IEEEmembership{Senior Member,~IEEE}%
\thanks{This work was supported in part by the Spanish and Catalan Governments
under grants TEC2015-69868-C2-2-R (ADVENTURE), TEC2014-59255-C3-1-R (ELISA) and
2014-SGR-1567.}
\thanks{* Contact author. D.~Gregoratti and X.~Mestre are with the Centre Tecnol\`ogic de Telecomunicacions de Catalunya (CTTC),
Parc Mediterrani de la Tecnologia, 08860~Castelldefels, Barcelona~(Spain),
e-mails: \{david.gregoratti, xavier.mestre\}@cttc.es. Part of the contents of
the paper was presented at IEEE SPAWC 2015.}}%

\title{Uplink FBMC/OQAM-based Multiple~Access~Channel: Distortion Analysis\\ under
Strong Frequency~Selectivity}

\renewcommand{\Re}{\operatorname{\mathbb{R}e}}
\renewcommand{\Im}{\operatorname{\mathbb{I}m}}
\newcommand{\odd}{\text{odd}}
\newcommand{\even}{\text{even}}
\newcommand{\bm}{\mathbf}
\DeclareMathOperator{\diag}{diag}
\DeclareMathOperator{\tr}{tr}
\newcommand{\cj}{\mathrm{j}}
\DeclareMathOperator{\EV}{\mathbb{E}}
\newcommand{\bs}{\boldsymbol}
\newcommand{\SNDR}{\mathit{SNDR}}
\newcommand{\SNR}{\mathit{SNR}}
\newcommand{\Mid}{\mathbf{I}}

\newtheorem{proposition}{Proposition}
\newtheorem{lemma}{Lemma}

\begin{document}

\maketitle
\thispagestyle{empty}

\begin{abstract}
This paper computes the distortion power at the receiver side of an
FBMC/OQAM-based OFDMA uplink channel under strong frequency selectivity and/or
user timing errors.  More precisely, it provides a distortion expression that is
valid for a wide class of prototype pulses (not necessarily
perfect-reconstruction ones) when the number of subcarriers is sufficiently
large. This result is a valuable instrument for analyzing how users interfere to
one another and to justify, formally, the common choice of placing an empty
guard band between adjacent users. Interestingly, the number of out-band
subcarriers contaminated by each user only depends on the prototype pulses and
not on the channel nor on the equalizer. To conclude, the distortion analysis
presented in this paper, together with some simulation results for a realistic
scenario, also provide convincing evidence that FBMC/OQAM-based OFDMA is
superior to classic circular-prefix OFDMA in the case of asynchronous users.
\end{abstract}

\begin{IEEEkeywords}
Filterbank, FBMC/OQAM, OFDMA, Multiple Access Channel, Strong Frequency
Selectivity.
\end{IEEEkeywords}

\section{Introduction}
Multicarrier techniques are based on the idea that a frequency-selective channel
can be split into a number of flat orthogonal subchannels. As a result,
sophisticated channel equalization schemes can be replaced by simple (typically
one-tap) per-subcarrier equalizers. For this reason, both wired (e.g., digital
subscriber line, powerline communications) and wireless (e.g., WiFi, WiMAX)
communications systems have long been employing multicarrier strategies in order
to control Intersymbol Interference (ISI) while limiting complexity
\cite{Goldsmith_Wireless}.

Recently, established communications standards like LTE have based also their
multiple access features on a multicarrier approach \cite{3GPP-TS36.211}.
Besides simple equalization, multicarrier transmission offers a flexible
mechanism to dynamically allocate variable portions of the spectrum to users,
according to their throughput requirements and the channel response.
Theoretically, it consists in a trivial assignment of subcarriers to users. In practice,
however, things may become substantially more complicated, depending on the synchronization
requirements between users and Base Station (BS) and, especially, among different
users.

In~\cite{Morelli_etal}, M.~Morelli et~al.\ analyze the synchronization problem
for the Orthogonal Frequency Division Multiple Access (OFDMA) scheme, which is
probably the most common multicarrier multiple-access strategy based
on an extension of the well-known Circular Prefix Orthogonal Frequency
Multiplexing (CP-OFDM) scheme. The authors highlight the fact that, similarly to
CP-OFDM, OFDMA suffers from poor spectrum containment and is hence very
sensitive to frequency offsets. The issue is exacerbated in the uplink, since
synchronism is needed among signals at the receiver side (and not at
the transmitters). Timing- and frequency-tracking algorithms are presented
in~\cite{Morelli_etal}, together with interference-cancellation procedures that
are required to remove residual interference. The resulting receiver is thus a
complex system and the inherent efficiency loss caused by the presence of the CP
is not justified anymore.

Filterbank Multicarrier (FBMC) modulation is an old technique (see,
e.g.,~\cite{Weinstein1971}) that is regaining popularity in the last few years as a
potential solution to the efficiency and synchronization limitations of CP-OFDM
and CP-OFDMA \cite{Vaidy2001, Farhang2011}. In FBMC, orthogonality among
subchannels is obtained by means of well-designed filters with low side lobes,
and is much less sensitive to frequency offsets as shown in, e.g., \cite{Roque2012}. Furthermore, no CP is needed,
thus improving the spectral efficiency. One should be aware, however, that one-tap
per-subcarrier equalizers are not ideal anymore (especially in highly frequency
selective channels) and more sophisticated solutions are often needed
\cite{Ihalainen2007,Waldhauser2008,Ndo2012,Mestre2013}. For this
reason, and because of FBMC architectures being more complex than their
CP-OFDM counterparts, FBMC is still not very popular in point-to-point
communications. On the other hand, the complexity gap cancels out (or possibly
reverses) in multiple-access scenarios, as discussed above (see, e.g., \cite{Saeedi_etal,
Mattera2015}). For instance, users do not need to be synchronized since the
timing at each subcarrier can be corrected separately.

\begin{figure}
\centering
\includegraphics{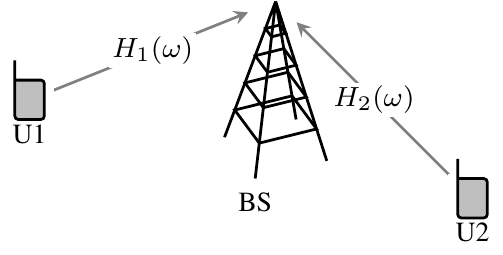}
\caption{An uplink frequency selective multiple-access channel with two users. }\label{fig:scenario}
\end{figure}

This paper considers a frequency selective multiple-access uplink channel as the
one depicted in Fig.~\ref{fig:scenario} and characterizes the distortion of the
received symbols assuming an FBMC/OQAM-based OFDMA scheme. Users are not
synchronized and their channels towards the BS are highly frequency selective.
No particular hypothesis is formulated about the FBMC prototype pulses, while a
single-tap per-subcarrier equalizer is implemented. We mentioned before that
this equalizer is suboptimal; however, a rigorous mathematical analysis of more
sophisticated receiver architectures would result in extremely complex
derivations. Moreover, single-tap equalizers are often used in real practical
systems due to their simplicity.

The resulting distortion is
thus the joint effect of suboptimal equalization and a ``poor'' filter
choice\footnote{In some application, system designers may decide to relax the
perfect reconstruction constraints as long as the resulting distortion is
negligible with respect to the equalization one and/or the noise level.}. As
opposed to other works, where an empirical approach is preferred
\cite{Saeedi_etal, Ihalainen2009}, the analysis below is based on a tight
approximation that accurately represents the received signal when the number
of subcarriers is large enough. This approximation is based on the results of
\cite{Mestre2013} and follows similar lines. It is worth remarking, however,
that the problem at hand presents some specific difficulties that cannot be
seen as simple extensions of \cite{Mestre2013}. Indeed, important properties
of the involved operations [e.g., of the Discrete Fourier Transform (DFT)] do
not hold when the spectrum is not considered in its integrity, but is split
among the different users.

A fine characterization of the distortion brings valuable help with the design of
an FBMC-based OFDMA system. Each user contributes not only to the distortion at
its assigned subcarriers (\emph{in-band distortion}), but also at other users'
subcarriers (\emph{out-band distortion}). Indeed, the interference caused by
frequency selective channels, inherent to FBMC schemes, is exacerbated in the
multiple user case since all users undergo different channels, whose combined
effect may be unpredictable. The purpose of this analysis is to confirm the
intuition that, with sharp prototype pulses, leaving one empty subcarrier as a
guard-band is sufficient \cite{Ihalainen2009, Mattera2015B}. Moreover, it
also applies to prototype pulses that are not so frequency selective, like those
used in double-dispersive channels and designed to optimize the time--frequency
localization of the waveform \cite{Siohan_etal,Roque2012,Siclet2006,Amini2015}.
Luckily, according to the results below, the number of
subcarriers affected by the out-band leakage only depends on the prototype
pulses at both sides of the FBMC links. Conversely, channel responses, equalizer
and synchronization misalignments only affect the distortion magnitude. This
means that no channel state information is needed to choose the prototype pulses
that minimize the leakage effect or to decide whether one or more empty
guard-bands are needed between users.

The rest of the paper is organized as follows. Section~\ref{sec:signal_model}
provides the model for the FBMC-based OFDMA channel under consideration while
recalling basic concepts of FBMC modulation. Then, the derivation and the
interpretation of the distortion approximation are reported in
Section~\ref{sec:distortion_analysis}. Finally,
Section~\ref{sec:numerical_analysis} contains some numerical assessment of the
results and Section~\ref{sec:conclusions} concludes the paper.

\subsubsection*{Notation}
Hereafter, lowercase (respectively, uppercase) boldface letters denote column
vectors (respectively, matrices). Occasionally, matrices that are functions of
other matrices are denoted by uppercase calligraphic letters. Superscripts
$(\cdot)^*$, $(\cdot)^T$ and $(\cdot)^H$ represent complex conjugate,
transpose and complex (Hermitian) transpose, respectively. For a generic matrix
$\bm A$, $[\bm A]_{k,l}$ is its $(k,l)$ entry. Also, borrowing from
Matlab$^\text{\textregistered}$ notation, $[\bm A]_{:,l}$ and $[\bm A]_{k,:}$ denote the
$l$-th column and the $k$-th row of $\bm A$, respectively. The entries of the
diagonal matrix $\diag\{\bm a\}$ (respectively, $\diag_{n=1,\dots,N}\{a_n\}$)
are the elements of vector $\bm a$ (respectively, of the sequence
$a_1,\dots,a_N$). $\tr A$ stands for the trace of matrix $\bm A$. $\Re\{\bm A\}$
and $\Im\{\bm A\}$ are the real and imaginary parts of $\bm A$, so that $\bm
A=\Re\{\bm A\} + \cj \Im\{\bm A\}$, with $\cj$ the imaginary unit. The
operators $\otimes$, $\odot$ and $\circledast$ stand for Kronecker product,
Hadamard (element-wise) product and row-wise convolution, respectively
(matrix dimension restrictions apply), while $\EV[\cdot]$ is the expected value. Symbol
$\bm 0_{m,n}$ (respectively, $\bm 1_{m,n}$) denotes an $m\times n$ matrix with
all entries equal to 0 (respectively, to 1). For simplicity, subscripts may be
removed (i.e., $\bm 0$ and $\bm 1$) from column vectors whose length can be
clearly determined by the context. Finally, $\bm I_k$ and $\bm J_k$ represent
the $k\times k$ identity and anti-identity (with one-valued entries only on the main
anti-diagonal) matrix, respectively.

\section{Signal Model}\label{sec:signal_model}

\begin{figure}
  \centering
  \includegraphics{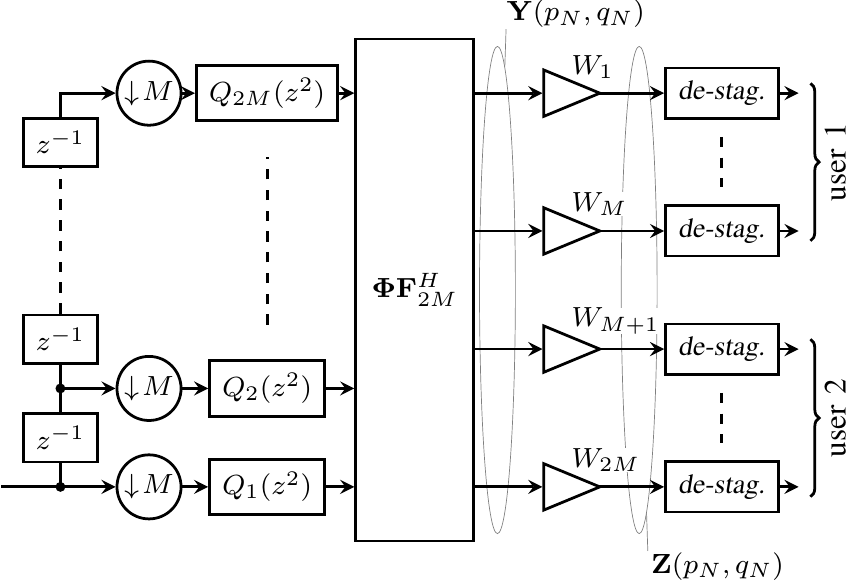}
  \caption{Block diagram of an FBMC/OQAM-based OFDMA receiver assuming that subcarriers are equally split
  between two users. Blocks labeled ``\textsl{de-stag.}'' represent the
  de-staggering operation (see, e.g.,  \cite{Mestre2013}). The coefficients of
  the polyphase filters $Q_m(z^{-2}) = \sum_{n=1}^\kappa
  q_N[m+2(n-1)M]z^{-2(n-1)}$ are given by the $m$-th row of matrix~$\bm Q$.
}\label{fig:scheme}
\end{figure}

In this paper we focus on an OFDMA channel obtained by means of an FBMC
modulation based on Offset Quadrature Amplitude Modulation (FBMC/OQAM, also
known as staggered modulated multitone). More precisely, we are interested in
the uplink channel where a common sink (e.g., a BS) receives data from
$K$ users (the case of two users is depicted in Fig.~\ref{fig:scenario}):
each user is assigned a subset of the available
$2M$ equally spaced subcarriers,
according to some allocation policy. Let $\mathcal{I}_k$
denote the set of subcarrier indices reserved for user $k$, so that
$\bigcup_{k=1}^K \mathcal{I}_k = \{1,\dots,2M\}$ and $\mathcal{I}_k \cap
\mathcal{I}_{k'} = \emptyset$ for all $k'\ne k$. Since channel and receiver are
linear systems, the received signal can be written as the sum of user
contributions. Then, for user $k$, let $\bm A_k = \bm B_k + \cj \bm C_k$ be the
$2M\times N_s$ matrix representing a block of $N_s$ multicarrier QAM symbols:
for all $n=1,\dots,N_s$ and all $m\in\mathcal{I}_k$,
entries $[\bm B_k]_{m,n}$ and $[\bm C_k]_{m,n}$ are independent (of one
another and across $k$, $m$ and $n$) real-valued bounded random variables with
zero mean and finite variance.
Conversely, for $m\notin\mathcal{I}_k$, the entries of $\bm A_k$ are identically
null.

\subsection{Memoryless Channel}

Let $p_N[n]$ and $q_N[n]$ be the real prototype pulses at the transmitter side and at
the receiver side, respectively. The length of both pulses is $N=2M\kappa$ taps,
where the overlapping factor $\kappa$ is an integer value. From the prototype
pulses we can build matrices
\begin{align}
\mathcal{R}(p_N,q_N) &= \begin{bmatrix}\bm P_1\circledast \bm J_M \bm Q_2 \\
\bm P_2\circledast \bm J_M \bm Q_1 \end{bmatrix} \label{eq:recR}\\
\mathcal{S}(p_N,q_N) &= \begin{bmatrix}\bm P_2\circledast \bm J_M \bm Q_2 \\
\bm P_1\circledast \bm J_M \bm Q_1 \end{bmatrix} \label{eq:recS}
\end{align}
where $\bm P_1$ and $\bm P_2$ gather the
top half and the bottom half rows, respectively, of matrix
$$
\bm P = \begin{bmatrix}\bm P_1\\ \bm P_2\end{bmatrix} = \begin{bmatrix}
p_N[1] & \cdots & p_N[2M(\kappa-1)+1]\\
\vdots &   & \vdots\\
p_N[2M] & \cdots & p_N[2M\kappa]\end{bmatrix}.
$$
Note that the $m$-th row of $\bm P$ contains the coefficients of the $m$-th
Type-I polyphase component of the prototype pulse $p_N[n]$. Equivalently, matrix
$\bm Q = {\begin{bmatrix}\bm Q_1^T & \bm Q_2^T\end{bmatrix}}^T$ is the polyphase
representation of the receiver prototype pulse $q_N[n]$.

For an ideal
(memoryless and noiseless) channel, the output of the analysis filterbank
corresponding to signal $\bm A_k$ (see Fig.~\ref{fig:scheme}) can be written as
$$
\bm Y_k(p_N,q_N) = \bm Y_k^{\even}(p_N,q_N)\otimes [1,0] + \bm
Y_k^{\odd}(p_N,q_N)\otimes [0,1]
$$
where
\begin{subequations}\label{eq:ideal_comp}
\begin{align}
\bm Y_k^\odd(p_N,q_N) = 2\bm\Phi\bm F_{2M}^H \Bigl([\bm F_{2M}\bm\Phi^*\bm B_k, \bm
0, \bm 0]\circledast\mathcal{R}(p_N,q_N)\Bigr)\nonumber\\
{}+2\bm\Phi\bm F_{2M}^H \Biggl(
\begin{bmatrix}\bm 0,\cj\bm G_2 \bm\Phi^*\bm C_k, \bm 0\\
\cj\bm G_1\bm\Phi^*\bm C_k,\bm 0, \bm 0\end{bmatrix}
\circledast\mathcal{S}(p_N,q_N)\Biggr),\\
\bm Y_k^\even(p_N,q_N) = 2\bm\Phi\bm F_{2M}^H \Bigl([\bm 0,\cj\bm
F_{2M}\bm\Phi^*\bm C_k, \bm 0]\circledast\mathcal{R}(p_N,q_N)\Bigr)\nonumber\\
{}+2\bm\Phi\bm F_{2M}^H \Biggl(
\begin{bmatrix}\bm 0,\bm G_2 \bm\Phi^*\bm B_k, \bm 0\\
\bm G_1\bm\Phi^*\bm B_k,\bm 0, \bm 0\end{bmatrix}
\circledast\mathcal{S}(p_N,q_N)\Biggr).
\end{align}
\end{subequations}
and where we have introduced the diagonal matrix $\bm \Phi =
\diag_{m=1,\dots,2M}\Bigl\{\exp\Bigl[-\cj \pi \frac{M+1}{2M}(m-1)\Bigr]
\Bigr\}$ and the $2M\times 2M$ Fourier matrix $\bm F_{2M}$ with entries $[\bm
F_{2M}]_{m,n} = (2M)^{-1/2} \exp\Bigl[\cj\frac{2\pi}{2M}(m-1)(n-1)\Bigr]$. The
upper (respectively, lower) $M$ rows of $\bm F_{2M}$ are denoted by $\bm G_1$
(respectively, $\bm G_2$), so that $\bm F_{2M}={\begin{bmatrix}\bm G_1^T & \bm
G_2^T\end{bmatrix}}^T$. Equation (\ref{eq:ideal_comp}) is the input--output
FBMC/OQAM signal model according to the efficient polyphase implementation
\cite{Siohan_etal, Mestre2013}. It is worth remarking that, even though based on
the polyphase formulation, the results below characterize the distortion for all the
equivalent implementations of the FBMC/OQAM architecture, namely the classical
transmultiplexer implementation with complex modulated prototype pulses
\cite{Siohan_etal}, the frequency-spreading formulation
\cite{Bellanger2012,Mattera2012} and, in some extent, the fast-convolution based
FBMC \cite{Renfors2014}.

As it can be evinced from (\ref{eq:ideal_comp}), there exists a complex
relationship between the transmitted symbols and the received ones. However, as
proven in \cite{Siohan_etal, Farhang, Mestre2013}, for instance, message recovery is
possible since
\begin{equation}\label{eq:estimate_id}
{[\bm A_k]}_{m,n} = \Re{[\bm Y_k^\odd(p_N,q_N)]}_{m,n+\kappa-1}
  +\cj\Im{[\bm Y_k^\even(p_N,q_N)]}_{m,n+\kappa}
\end{equation}
whenever the prototype pulses meet the Perfect Reconstruction (PR) constraints
\begin{subequations}\label{eq:RC}
\begin{align}
  \bm U^+ \mathcal{R}(p_N,q_N) &= \mathbb{I} \label{eq:RC1} \\
  \bm U^- \mathcal{S}(p_N,q_N) &= \bm 0_{2M\times(2\kappa-1)}. \label{eq:RC2}
\end{align}
\end{subequations}
Matrices $\bm U^+$ and $\bm U^-$ are defined as
\begin{equation}\label{eq:Udef}
  \bm U^\pm = \bm I_2 \otimes (\bm I_M \pm \bm J_M)
\end{equation}
while $\mathbb{I}=[\bm 0_{2M\times (\kappa-1)}, \bm 1, \bm 0_{2M\times
(\kappa-1)}].$

\subsection{Frequency-Selective Channel}

As mentioned above, multicarrier modulations find their main application when the
communication channel is frequency selective. In such a situation, however, the
memoryless model in (\ref{eq:ideal_comp}) does not hold anymore. Indeed, due to the
delay spread, the received signal can be seen as a weighted combination of a
number of delayed replicas. The delay and the weight of each replica depend on
the channel impulse response. Note that timing errors between transmitter and
receiver are covered by this model, since they are equivalent to a phase shift
in the channel frequency response. Conversely, frequency offsets need a more
sophisticated analysis that is not treated in this paper in order to avoid
further complexity. In other words, we are assuming that channel estimations are
refreshed often enough to neglect the Doppler effect.

Under a frequency selectivity assumption, a more useful approximation for the
output of the analysis filterbank in Fig.~\ref{fig:scheme} is given by
\cite{Mestre2013}
\begin{multline}\label{eq:freqsel}
\bm Z_k^{(*)}(p_N,q_N) = \bm W \bm \Lambda_{H_k} \bm Y_k^{(*)}(p_N,q_N)
-\frac{\cj}{2M}\bm W \bm \Lambda_{H_k^{(1)}} \bm
Y_k^{(*)}\Bigl(p_N,q_N^{(1)}\Bigr)\\
{}-\frac{1}{8M^2}\bm W \bm \Lambda_{H_k^{(2)}} \bm
Y_k^{(*)}\Bigl(p_N,q_N^{(2)}\Bigr) +
\bm o\bigl(M^{-2}\bigr)
\end{multline}
where $(*)\in\{\odd,\even\}$ and where\begin{itemize}
\item the diagonal matrix $\bm W$ collects the $2M$ coefficients of a channel
equalizer with a single tap per subcarrier;
\item the $m$-th entry of the
diagonal matrix $\bm \Lambda_{H_k^{(r)}}$ is $H_k^{(r)}(\omega_m)$, that is the
value of the $r$-th derivative of the channel frequency response
$H_k(\omega)$ computed at $\omega=\omega_m$, the central frequency of subcarrier
$m$;
\item the symbols $\bm Y_k^{(*)}\Bigl(p_N,q_N^{(r)}\Bigr)$ are built as in
(\ref{eq:ideal_comp}) after substituting the receiver prototype pulse $q_N$ with its
\mbox{$r$-th} derivative $q_N^{(r)}$ (see below);
\item $\bm o\bigl(M^{-2}\bigr)$ represents a matrix of appropriate size whose
entries decay faster than $M^{-2}$ as $M\to+\infty$.
\end{itemize}

\begin{figure}[t!]
  \centering
  \includegraphics{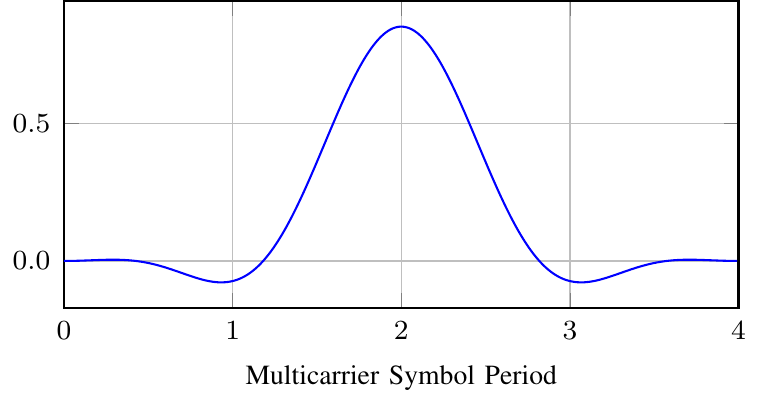}
  \caption{Prototype pulse defined by the EU-funded project PHYDYAS
  \cite{TR:PHYDYAS-D5.1} with overlapping factor $\kappa=4$: it is a good approximation of a prototype pulse
  satisfying the required assumptions.}\label{fig:pulse}
\end{figure}

The approximation in (\ref{eq:freqsel}) is derived for an asymptotically large
number $2M$ of subcarriers under the following assumptions.
\begin{description}
\item[\slshape\bfseries AS1] The coefficients of $p_N[n]$ and $q_N[n]$ are uniformly
bounded for $M\to+\infty$.
\item[\slshape\bfseries AS2] The pulse $q_N[n]=q_N^{(0)}[n]$ and its derivatives $q_N^{(r)}[n]$ are
obtained according to
$$
q_N^{(r)}[n] = T_s^r q^{(r)}\biggl(\biggl(n-\frac{N+1}{2}\biggr)
\frac{T_s}{2M}\biggr),
$$
where $T_s$ is the multicarrier symbol period in time units and where $q(t)$ is a twice continuously
differentiable real function defined over $[-\kappa T_s/2,\kappa T_s/2]$.
Moreover, the function and its first two derivatives null out at the boundary
points of the domain, that is $q(\pm \kappa T_s/2) = q^{(1)}(\pm \kappa T_s/2) =
q^{(2)}(\pm \kappa T_s/2) = 0$. An example is depicted in Fig.~\ref{fig:pulse}.
\end{description}
Recall that the subcarrier bandwidth, and hence the multicarrier symbol
duration, are kept constant. This means that, by increasing the number of
subcarriers, we are considering a wider spectrum (or, equivalently, a higher
sampling frequency). This assumption may seem not very practical; nevertheless,
numerical results will show how asymptotic expressions finely approximate real
systems with reasonable total bandwidth and number of subcarriers.

Even though the expansion in (\ref{eq:freqsel}) can be easily extended to any
order higher than two, the extra terms would not bring any significant
contributions to the analysis below. Numerical results in
Section~\ref{sec:numerical_analysis} will confirm this statement and the fact
that (\ref{eq:freqsel}) is representative of practical systems with a
sufficiently large number of subcarriers. Similarly, more sophisticated
equalizers (e.g., single-tap minimum mean square error or even more complex
schemes like those described in \cite{Ihalainen2007,Waldhauser2008,Ndo2012,Mestre2013})
could be considered and included in our study. However, this choice
would make the mathematical derivations extremely complicated and hard to
follow.

\section{Distortion Analysis}\label{sec:distortion_analysis}

\subsection{Preliminary considerations}

Before delving into the analysis of the distortion affecting the considered
system, we need to clarify how distortion is defined. Let us focus on
subcarrier $m$ assigned to user $\ell$, i.e.\ $m\in\mathcal{I}_\ell$. Then, ideally,
the symbols received on this subcarrier would be $W_m H_\ell(\omega_m) [\bm A_\ell]_{m,n}$,
where $W_m$ [respectively, $H_\ell(\omega_m)$] is the $m$-th entry of the equalizer matrix $\bm W$
(respectively, of the channel matrix $\bm \Lambda_{H_\ell}$) and where
$n=1,\dots,N_s$. Note that we consider the general case where the equalizer does
not necessarily invert the channel, i.e.\ we do not necessarily fix $W_m =
[H_\ell(\omega_m)]^{-1}$. Nevertheless, since both channel and equalizer
are assumed to be known, symbols $[\bm A_\ell]_{m,n}$ can be readily recovered.

After the de-staggering operation, the receiver output
corresponding to subcarrier $m$ is
[cf.\ (\ref{eq:estimate_id})]
\begin{equation}\label{eq:a_estimate}
\tilde{a}_{m,n} = \sum_{k=1}^K \Bigl(\Re{[\bm Z_k^\odd(p_N,q_N)]}_{m,n+\kappa-1}
+\cj\Im{[\bm Z_k^\even(p_N,q_N)]}_{m,n+\kappa}\Bigr).
\end{equation}
It is important to remark that the contributions of all users must be taken into
account. Indeed, similarly to the single-user case (see, e.g.,
\cite{Siohan_etal, Farhang, Mestre2013}), the inherent structure of FBMC modulations [see
(\ref{eq:ideal_comp})] and the channel delay spread [see
(\ref{eq:freqsel})] produce Intercarrier Interference (ICI), regardless of
whether the prototype pulses satisfy the PR conditions. The ICI due to each
user is not confined within the subcarriers assigned to the user itself (in-band
distortion), but leaks into other users' subcarriers (out-band distortion), as
proven later.

For the symbol estimates $\tilde{a}_{m,n}$ in (\ref{eq:estimate_id}), we define
the distortion at subcarrier $m\in \mathcal{I}_\ell$ as
$$
P_e(m) = \EV \Bigl|\tilde{a}_{m,n} - W_m H_\ell(\omega_m)[\bm
A_\ell]_{m,n}\Bigr|^2.
$$
Recalling that $[\bm A_k]_{m,n}=0$ for all $k\ne \ell$ (since $m\in
\mathcal{I}_\ell$), we can further write
\begin{align*}
P_e(m) &= \EV \Bigl|\tilde{a}_{m,n} - W_m \sum_{k=1}^K H_k(\omega_m) [\bm
A_k]_{m,n}\Bigr|^2 \\
&= 2\EV\sum_{k=1}^K \biggl|\Re[\bm Z_k^{\odd}(p_N,q_N)]_{m,n+\kappa-1}
- W_m H_k(\omega_m) [\bm B_k]_{m,n}\biggr|^2.
\end{align*}
The second equality holds after assuming that the entries of the real part (i.e.\ $\bm
B_k$) and of the imaginary part (i.e.\ $\bm C_k$) of $\bm A_k$ are independent
and identically distributed, apart from being independent across the user index
$k$. If this were not the case, we would need to consider also the difference
between $\Im[\bm Z_k^{\even}(p_N,q_N)]_{m,n+\kappa}$ and $W_m H_k(\omega_m) [\bm
C_k]_{m,n}$. Such extension is straightforward and not developed here.

To simplify notation, let us write $z_{k,m}^\odd = [\bm Z_k^\odd]_{m,n}$ and
$b_{k,m}=[\bm B_k]_{m,n}$. Then, the last equation tells us that each user $k$
contributes to the distortion at subcarrier $m$ with a term
\begin{equation}\label{eq:user_contr}
P_{e,k}(m) = 2\EV\Bigl|\Re[z_{k,m}^\odd]-W_m H_k(\omega_m) b_{k,m}\Bigr|^2.
\end{equation}
Note that all user contributions take the same form and it does not matter
whether subcarrier $m$ is assigned to user $k$ (that is, $m\in \mathcal{I}_k$
and $b_{k,m}\ne 0$) or to another user (i.e., $m\notin \mathcal{I}_k$ and
$b_{k,m} = 0$). In other words, according to the value of $m$, (\ref{eq:user_contr})
represents either the in-band or the out-band distortion due to user $k$.

\subsection{Symmetric PR-compliant Prototype Pulses}
By plugging (\ref{eq:freqsel}) into (\ref{eq:user_contr}), one can compute the
distortion corresponding to a prototype pulse with generic time and frequency
responses, as long as it satisfies assumptions \textbf{\textsl{AS1}} and
\textbf{\textsl{AS2}}.
However, as shown in Appendix~\ref{sec:proof_dist}, the
resulting expression is cumbersome and offers no hints for
interpretation. For this reason, hereafter we
look for insight into the special simple case where the prototype pulse is
symmetric and compliant with the PR conditions in (\ref{eq:RC}). Note that
prototype pulses meeting all these requirements can be actually designed, as
shown in \cite{Mestre2013}.

\begin{proposition}\label{prop:dist_terms}
Consider the FBMC/OQAM-based OFDMA channel described above. Apart from
satisfying assumptions \textbf{\textsl{AS1}} and \textbf{\textsl{AS2}},
choose two prototype pulses $p_N[n]$ and $q_N[n]$ that meet PR conditions
(\ref{eq:RC}) and are symmetric (e.g., $p_N[n]=p_N[N-n+1]$). Then, the
contribution of user~$k$ to the distortion at subcarrier $m$ can be written as
\begin{equation}\label{eq:dist_formula}
  P_{e,k}(m) = 2\biggl[\eta_{0,0}(m) + \frac{2}{2M}\eta_{0,1}(m) +
\frac{1}{4M^2}\eta_{0,2}(m) + \frac{1}{4M^2}\eta_{1,1}(m)\biggr] + o\bigl(M^{-2}\bigr)
\end{equation}
where terms $\eta_{x,y}(m)$ are given by
\begin{subequations}\label{eq:dist_terms}
\begin{align}
\eta_{0,0}(m) &= 2 \Im^2[W_m H_k(\omega_m)]\Re\tr\biggl[\bm U^- \mathcal F_k(m) \mathcal{X}^{\mathcal R-\frac{1}{2}\mathbb{I}}_{(p,q,p,q)}
+ \bm U^+ \mathcal F_k(m) \mathcal{X}^{\mathcal S}_{(p,q,p,q)}\biggr]\\
\eta_{0,1}(m) &= -2 \Im[W_m H_k(\omega_m)] \Im\Bigl[W_m
H_k^{(1)}(\omega_m)\Bigr]\nonumber\\&\qquad{}\times \Im \tr
\biggl[\bm U^+\mathcal F_k(m) \mathcal{X}^{\mathcal R-\frac{1}{2}\mathbb{I},\mathcal R}_{(p,q,p,q')}
+ \bm U^- \mathcal F_k(m) \mathcal{X}^{\mathcal S}_{(p,q,p,q')}\biggr]\\
\eta_{0,2}(m) &= -2 \Im[W_m H_k(\omega_m)] \Im\Bigl[W_m H_k^{(2)}(\omega_m)\Bigr]\nonumber\\&\qquad{}\times \Re \tr
\biggl[\bm U^-\mathcal F_k(m) \mathcal{X}^{\mathcal R-\frac{1}{2}\mathbb{I},\mathcal R}_{(p,q,p,q'')}
+ \bm U^+ \mathcal F_k(m) \mathcal{X}^{\mathcal S}_{(p,q,p,q'')}\biggr]\\
\eta_{1,1}(m) &= 2 \Im^2\Bigl[W_m H_k^{(1)}(\omega_m)\Bigr]\Re\tr\biggl[\bm U^+ \mathcal F_k(m) \mathcal{X}^{\mathcal R}_{(p,q',p,q')}
+ \bm U^- \mathcal F_k(m) \mathcal{X}^{\mathcal S}_{(p,q',p,q')}\biggr]\nonumber\\
&\qquad{}+ 2 \Re^2\Bigl[W_m H_k^{(1)}(\omega_m)\Bigr]\Re\tr\biggl[\bm U^- \mathcal F_k(m) \mathcal{X}^{\mathcal R}_{(p,q',p,q')}
+ \bm U^+ \mathcal F_k(m) \mathcal{X}^{\mathcal S}_{(p,q',p,q')}\biggr].
\end{align}
\end{subequations}
The
new matrices used in the definition of terms $\eta_{x,y}(m)$ are
\begin{align}
&\mathcal{F}_k(m) = \bm F_{2M}^H \EV\bigl[\bm b_k\bm b_k^H\bigr]\bm F_{2M} \odot
\bm f(m) \bm f^H(m),\nonumber\\
&\mathcal{X}^{\mathcal R}_{(p,q',p,q')} =
\mathcal{R}(p_N,q_N')\mathcal{R}^T(p_N,q_N'),\label{eq:Wr}\\
&\mathcal{X}^{\mathcal R-\frac{1}{2}\mathbb{I}}_{(p,q,p,q)} =
\Bigl(\mathcal{R}(p_N,q_N)-\frac{1}{2}\mathbb{I}\Bigr){\Bigl(\mathcal{R}(p_N,q_N)-\frac{1}{2}\mathbb{I}\Bigr)}^T,\nonumber\\
& \mathcal{X}^{\mathcal R-\frac{1}{2}\mathbb{I}, \mathcal R}_{(p,q,p,q^{(r)})} =
\Bigl(\mathcal{R}(p_N,q_N)-\frac{1}{2}\mathbb{I}\Bigr)\mathcal{R}^T(p_N,q_N^{(r)}),\nonumber\\
& \mathcal{X}^{\mathcal S}_{(p,q^{(r)},p,q^{(s)})} = \begin{bmatrix}
\mathcal S_1(p_N,q_N^{(r)})\mathcal S_1^T(p_N,q_N^{(s)}) & [0, \mathcal
S_1(p_N,q_N^{(r)})]{[\mathcal S_2(p_N,q_N^{(s)}),0]}^T\\
[\mathcal S_2(p_N,q_N^{(r)}),0]{[0, \mathcal S_1(p_N,q_N^{(s)})]}^T & \mathcal
S_2(p_N,q_N^{(r)})\mathcal S_2^T(p_N,q_N^{(s)})
\end{bmatrix} \label{eq:Xs}
\end{align}
where $\bm b_k$ is a generic column of $\bm B_k$ and $\bm f(m)$ is the $m$-th
column of $\bm F_{2M}$.
\end{proposition}
\begin{IEEEproof}
  The distortion expression for the considered special case is derived in
  Appendix~\ref{sec:proof_dist} as a particularization of the general one.
\end{IEEEproof}

When comparing (\ref{eq:dist_formula}) with \cite[Eq. (35)]{Mestre2013}, i.e.\
with the distortion expression for the single-user case\footnote{More properties on the
relationship between (\ref{eq:dist_formula}) and \cite[Eq. (35)]{Mestre2013} are
given in Appendix~\ref{sec:proof_dist}.}, one readily sees an important difference: as we increase
the number of subcarriers, the distortion decays as $M^{-2}$ in the single-user
case whereas terms of order $O(M^{-1})$ and $O(1)$ appear in the multi-user
case.  Fortunately, as explained hereafter, cross-user interference is
concentrated only at the users' boundaries. Thus, as we increase the total
number of subcarriers $2M$, the fraction of interfered spectrum becomes
proportionally smaller even though the interference magnitude does not fade out.

\subsubsection{Interpretation}\label{sssec:interpretation}
Even in the special case of Proposition~\ref{prop:dist_terms}, the
interpretation of the distortion expression in (\ref{eq:dist_formula}) is not
straightforward. We will now try to guide the reader through this task.

A closer look at (\ref{eq:dist_terms}) reveals that functions $\eta_{x,y}(m)$
are built from two types of factors. The first type of factors depends on the
equalizer and on the channel of the considered user. It is straightforward to
notice that, for the classic equalizer $W_m=H_k^{-1}(\omega_m)$, we have
$\Im[W_m H_k(\omega_m)]=0$ for $m\in\mathcal{I}_k$ and, thus, functions
$\eta_{0,0}(m)$, $\eta_{0,1}(m)$ and $\eta_{0,2}(m)$ bring no contribution to
the in-band distortion, but only to the out-band one, where the equalizer is
tuned on a different user's channel. Conversely, $\eta_{1,1}(m)$ only depends on
the first derivative of the channel response and, thus, it contributes to the
in-band distortion (compare also with \cite[Eq. (35)]{Mestre2013}). It is also
worth remarking that these factors show a strong dependence on the derivatives of the channel
frequency response. We have thus another evidence that it is generally incorrect
to assume that subcarrier channels are frequency flat.

\begin{figure}
  \centering
  \includegraphics{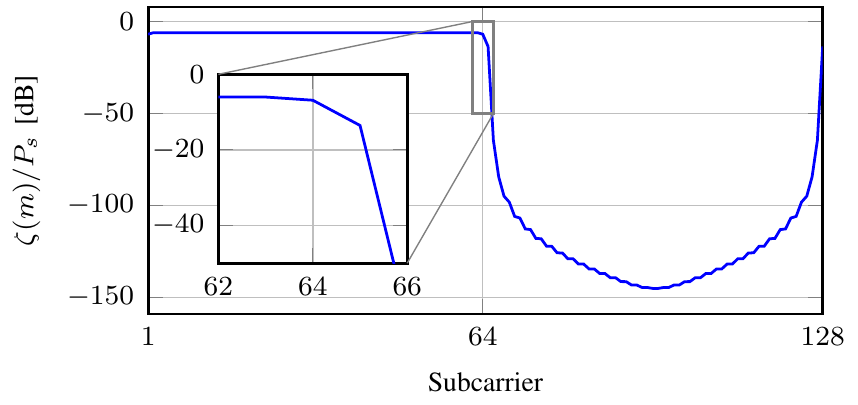}
  \caption{$\zeta(m)=\Re\tr\Bigl[\bm U^- \mathcal F_k(m)
      \mathcal{X}^{\mathcal R-\frac{1}{2}\mathbb{I}}_{(p,q,p,q)} + \bm U^+
  \mathcal F_k(m) \mathcal{X}^{\mathcal S}_{(p,q,p,q)}\Bigr]$ for a user
transmitting on subcarriers 1 to 64 with power $P_s$. The prototype pulse of
Fig.~\ref{fig:pulse} is employed at both the transmitter and the receiver side.}\label{fig:eta00}
\end{figure}

The second type of factors, on the other hand, corresponds to system design
variables, namely the prototype pulses, the user's assigned subcarriers and its
transmit power. Fig.~\ref{fig:eta00} depicts the behavior of the design factor
of $\eta_{0,0}(m)$, namely
$$
\zeta(m)=\Re\tr\Bigl[\bm U^- \mathcal F_k(m)
\mathcal{X}^{\mathcal R-\frac{1}{2}\mathbb{I}}_{(p,q,p,q)} + \bm U^+ \mathcal
F_k(m) \mathcal{X}^{\mathcal S}_{(p,q,p,q)}\Bigr]
$$
for a user transmitting on
subcarriers 1--64 with power $P_s$ and where the prototype pulse of
Fig.~\ref{fig:pulse} is used on both sides. As we can see, factor $\zeta(m)$ is
almost constant on the user subcarriers $\mathcal{I}_k=\{1,\dots,64\}$, while it
decays abruptly outside~$\mathcal{I}_k$. More specifically, the loss is about
7~dB at the first out-band subcarrier (i.e., number 65 and 128) and 58~dB at the
second out-band subcarrier (i.e., number 66 and 127). In other words, the
out-band interference is negligible except for the first subcarrier on both
sides of the user's assigned spectrum.

\begin{figure}
  \centering
  \includegraphics{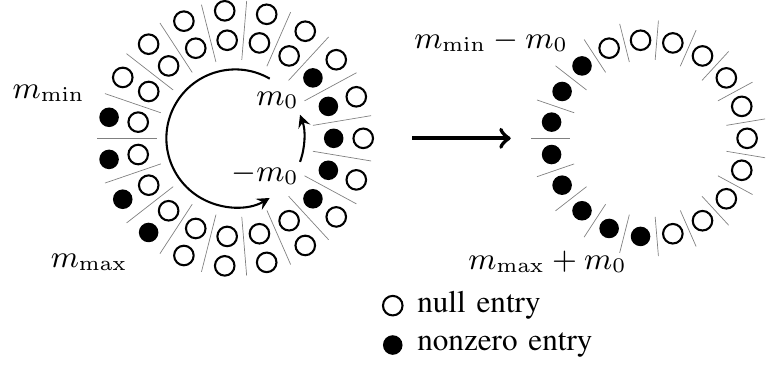}
  \caption{Circular cross-correlation of two vectors with few nonzero entries.
    Recall that $-m_0\equiv 2M-m_0$ modulo $2M$.}\label{fig:circ_corr}
\end{figure}

The behavior of $\zeta(m)$ is closely related to the frequency response
of the prototype pulses employed in the FBMC system. To
see this, consider the case where the overlapping factor is set to $\kappa=1$
(all other constraints, namely \textsl{\textbf{AS1, AS2,}} PR and symmetry,
still hold). Also, let $p_N[n]=q_N[n]$, that is the same pulse is used on both
sides. Then, matrix $\mathcal{R}(p_N, q_N)$ in (\ref{eq:recR}) is actually a
vector, namely
$$
\mathcal{R}(p_{2M},p_{2M}) = \bm r = \begin{bmatrix} p_{2M}^2[1] & \cdots &
p_{2M}^2[2M]\end{bmatrix}^T.
$$
Denote by $\bs \rho=[\rho_1\,\cdots\,\rho_{2M}]^T$ the DFT of $\bm r$, that is
$\bs \rho = \bm F_{2M}^H \bm r$. Simple algebra allows us to rewrite the first
term of $\zeta(m)$ as
\begin{equation}\label{eq:circ_corr}
\Re\tr\Bigl[\bm U^- \mathcal F_k(m)
\mathcal{X}^{\mathcal R-\frac{1}{2}\mathbb{I}}_{(p,q,p,q)}\Bigr] = 2\sum_{n=1}^M |\rho_{2n}|^2\beta_{k,2n-m+1}
\end{equation}
where $\beta_{k,n} = \EV[b_{k,n}^2]/(2M)$.
For well designed prototype pulses,
we will now prove that the right-hand
side of (\ref{eq:circ_corr}) is identically null for most $m\notin\mathcal{I}_k$.
Indeed, without loss of generality, we can assume that
$\mathcal{I}_k=\{m_{\min},\dots,m_{\max}\}$ and, thus, $\beta_{k,n}\ne 0
\Leftrightarrow m_{\min}\le n\le m_{\max}$. On the other hand, if
the prototype pulses are low-pass filters, the number of nonzero Fourier
coefficients is limited. More
specifically, there exists an integer value $m_0$ (typically
$m_0\approx2\kappa$, see, e.g., \cite{Mirabbasi_Martin}) such that
$\rho_n\ne 0$ if and only if $1\le n
\le m_0+1$ or $2M-m_0+1\le n \le 2M$ (recall that, since $\bm r$ is real,
$\rho_n = \rho_{2M-n+2}$ for $n=2,\dots,2M$). Then, a careful
inspection of the indices (Fig.~\ref{fig:circ_corr} may help in this purpose)
shows that $\zeta(m)$ is not zero if and only if\footnote{Index algebra is
modulo $2M$.} $m_{\min}-m_0 \le m \le m_{\max}+m_0$. Similar reasoning holds for
the second term of $\zeta(m)$ and for all the other distortion terms
$\eta_{x,y}(m)$. Conversely, by a similar reasoning, it is readily seen that the
distortion terms $\eta_{x,y}(m)$ may be significantly different than zero for
most $m$ if the spectrum containment property of the prototype pulses is not as
strict and the number of nonzero Fourier coefficients is high. In this category
we find pulses for time-limited orthogonal multicarrier modulation schemes
\cite{Li1995} and some pulses maximizing the time--frequency localization of the
signal \cite{Siohan_etal,Roque2012,Siclet2006,Amini2015}.

Summarizing the above discussion, an accurate design of the prototype pulses
permits the out-band distortion to be confined within few subbands at the
boundaries of the user of interest. A proper equalization (e.g.\ zero forcing), conversely, reduces
the in-band distortion, since terms $\eta_{0,0}(m)$, $\eta_{0,1}(m)$ and
$\eta_{0,2}(m)$ can be easily canceled out. Unfortunately, with the considered
single-tap-per-subcarrier equalizer, no degrees of freedom are left to further
minimize $\eta_{1,1}(m)$, which depends on the
first derivative of the channel frequency response.

\section{Numerical Analysis}\label{sec:numerical_analysis}
In this section, the theoretical results above are verified against numerical
ones obtained by simulating a realistic FBMC-based OFDMA channel. More
specifically, we consider a scenario with $2M=128$ subcarriers, spaced by
15~kHz. The resulting total bandwidth of 1.920 MHz is thus compliant with the
LTE standard \cite{3GPP-TS36.211}. The overlapping factor is set to $\kappa=4$ and
the prototype pulse proposed by the EU-funded project PHYDYAS \cite{TR:PHYDYAS-D5.1}, depicted in
Fig.~\ref{fig:pulse}, is used at both the transmitter side and the receiver
side. Note that this pulse is not PR-compliant and, thus, the full distortion
expression of Appendix~\ref{sec:proof_dist} shall be used.
We also assume that two users are accessing the channel and that each of them
is assigned 64 subcarriers (subcarrier 1 to 64 to User~1 and subcarrier 65 to
128 to User 2). The equalizer is tuned according to the zero-forcing principle on each
user's subband, i.e.\ $W_m=H_k^{-1}(\omega_m)$ for all $m\in\mathcal{I}_k$. All
subcarriers are subject to a white Gaussian noise process with variance
$\sigma^2$. Finally, the transmitted symbols are uniformly drawn from a 4QAM constellation
with power~$P_s$.

\subsection{Theory Assessment}
\begin{figure}
  \centering
  \includegraphics{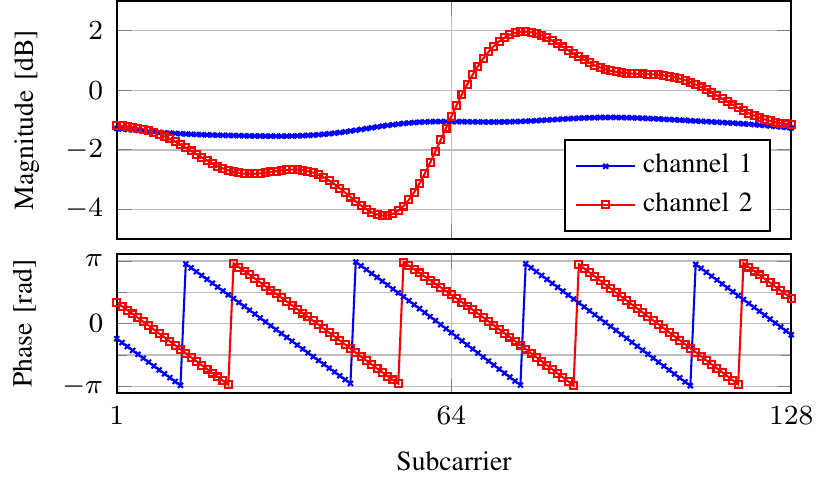}
  \caption{Two example realizations of EPA channel.}\label{fig:channelsEPA}
\end{figure}

\begin{figure}
  \centering
  \includegraphics{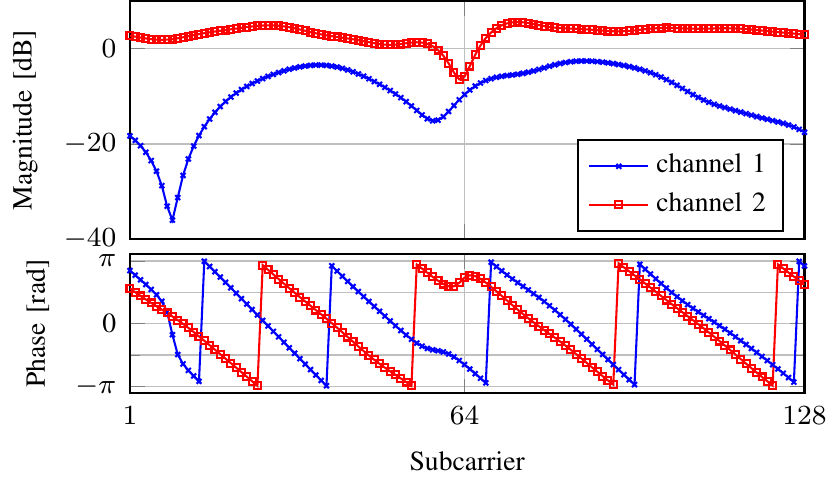}
  \caption{Two example realizations of EVA channel.}\label{fig:channelsEVA}
\end{figure}

\begin{figure}
  \centering
  \begin{tabular}{rr}
    \includegraphics{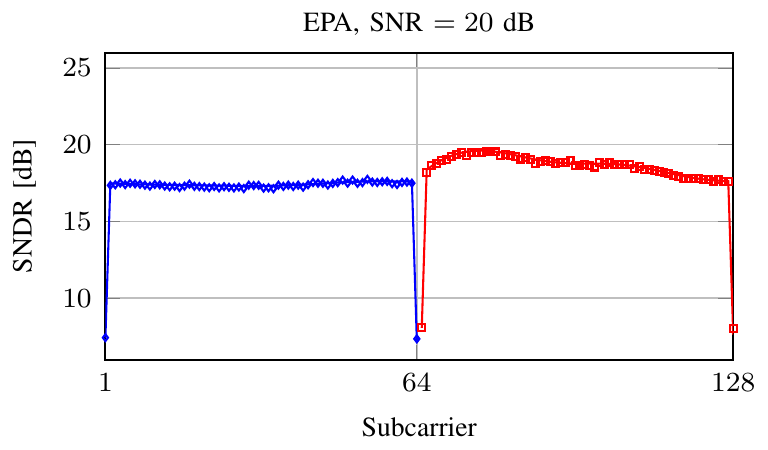}&
    \includegraphics{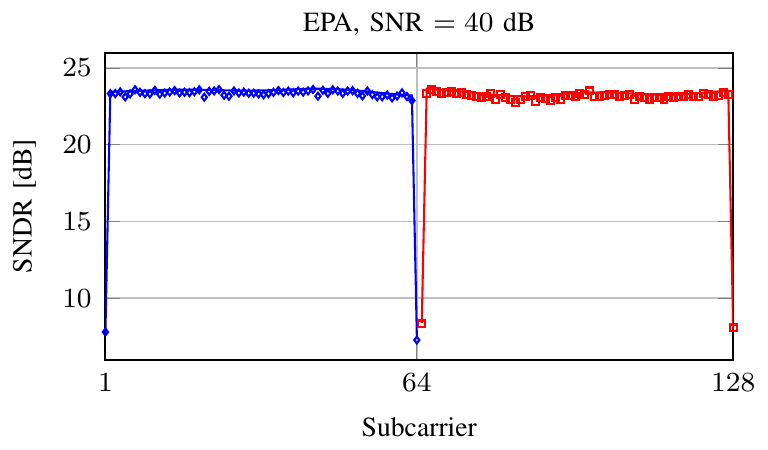}\\
    \includegraphics{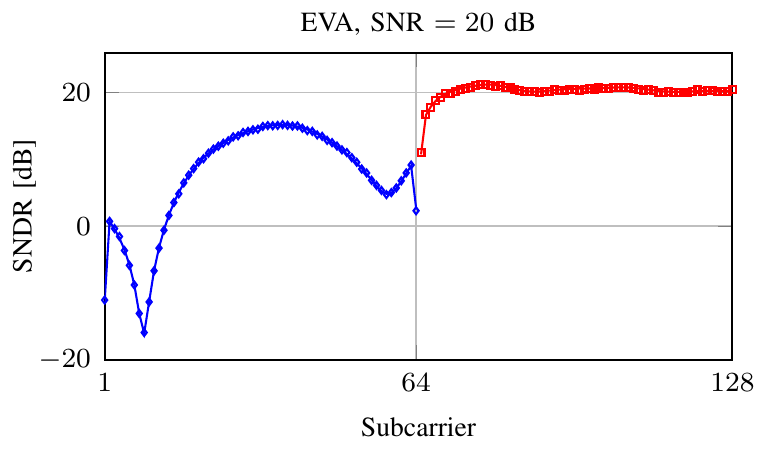}&
    \includegraphics{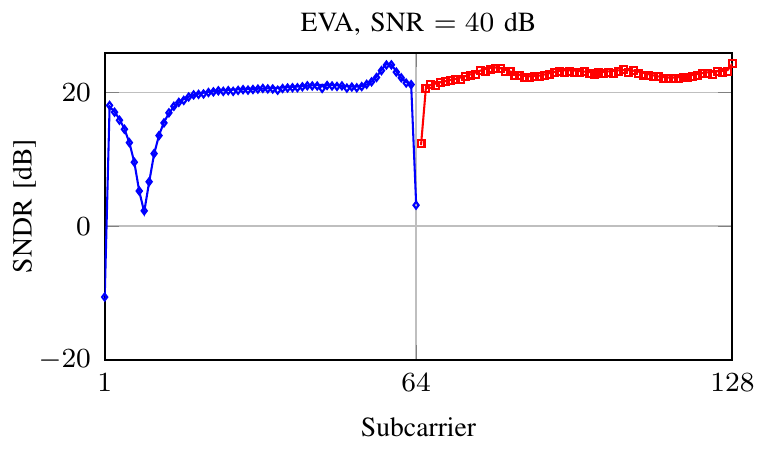}
  \end{tabular}
\caption{Theoretical (line) and empirical (markers) SNDR for the two-user case.
The ratio $P_s/\sigma^2$ is set to 20 dB (left) and 40 dB (right). Reported
examples correspond to EPA channel model (top) and EVA channel model (bottom).}\label{fig:SNDR}
\end{figure}

In order to test the above results with different degrees of
channel frequency selectivity, we assume that the channels of the two users 
follow either the ITU Extended Pedestrian~A (EPA) model or the ITU
Extended Vehicular A (EVA) model~\cite{3GPP-TR36.803}. More specifically, for the channel instances depicted
in Fig.~\ref{fig:channelsEPA} and Fig.~\ref{fig:channelsEVA}, we compute the
subcarrier Signal-to-Noise-and-Distortion Ratio (SNDR) according to
$$
\SNDR(m) = \frac{P_s}{P_e(m) + P_w(m)}
$$
where $P_e(m)=\sum_{k=1}^2P_{e,k}(m)$ is the total distortion, sum of users'
contributions as in
(\ref{eq:dist_formula}), and $P_w(m) = \sigma^2{|W_m|}^2
M^{-1} \sum_{n=1}^{N}q_N^2[n]$ is the noise power at the output of the
filterbank. For both channel cases and for a Signal-to-Noise Ratio
($\SNR=P_s/\sigma^2$) of either 20~dB or 40~dB, the results are depicted in
Fig.~\ref{fig:SNDR} and compared to the empirical values obtained by averaging
over 2000 multicarrier symbols. As one can observe, the two curves match perfectly.

\begin{figure}
  \centering
  \begin{tabular}{r}
    \mbox{\includegraphics{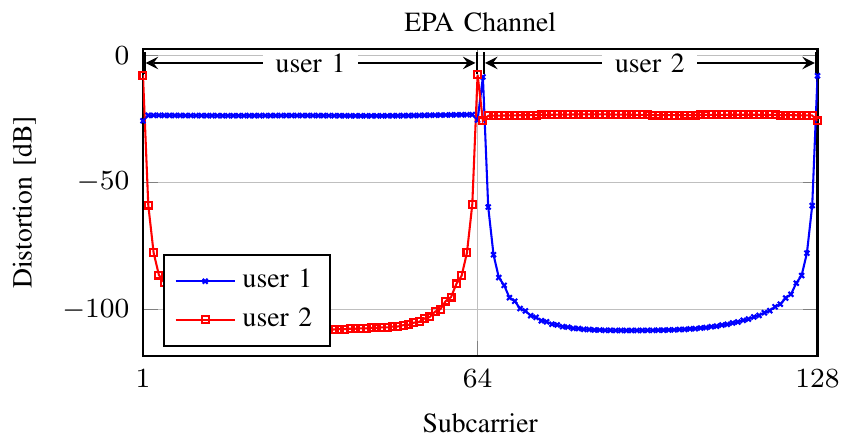}}\\%
    \mbox{\includegraphics{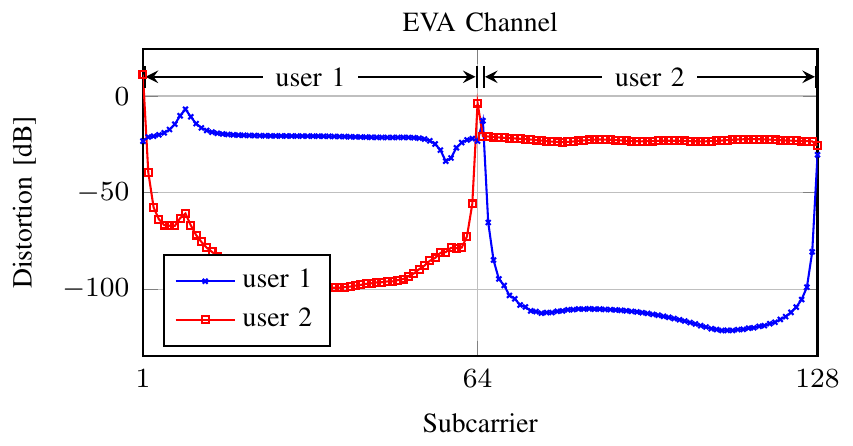}}%
  \end{tabular}
  \caption{Normalized distortion power for the two users. Example of EPA channel model (top)
  and of EVA channel model (bottom).}\label{fig:distortion}
\end{figure}

It is interesting to note that, when increasing the transmit power from 20~dB to
40~dB over the noise level, the SNDR improves by a mere 6~dB and saturates at
around 23~dB. This is indeed the inverse of the distortion level $P_e(m)/P_s$,
as one can appreciate from Fig.~\ref{fig:distortion}. The curves of
Fig.~\ref{fig:distortion} also confirm that the in-band distortion (the one
generated by a user in its own subband) is the main cause of signal degradation
at high SNR. The out-band distortion, on the other hand, is significant only in
the first subcarrier outside the user's subband, as evinced from the peaks in
Fig.~\ref{fig:distortion} (and from the dips in Fig.~\ref{fig:SNDR}). This could
be predicted according to Section~\ref{sssec:interpretation}, keeping in mind
that the Fourier coefficients of the prototype pulse in use are those reported
in Table~\ref{tab:pulseFcoef} (only $M$ coefficients are needed since the pulse is real-valued).

\begin{table}
\renewcommand{\arraystretch}{1.3}
\caption{Pulse Fourier Coefficients}\label{tab:pulseFcoef}
\centering
\begin{tabular}{c||c|c|c|c|c}
\hline
\textbf{Index} & 1 & 2 & 3 & 4 & 5--256\\\hline
\textbf{Magnitude} & 4.000 & 3.888 & 2.828 & 0.940 & 0\\\hline
\end{tabular}
\end{table}

\subsection{Delay Channels}\label{ssec:delay}
In order to confirm the interest for FBMC/OQAM as an enabler technology for
OFDMA, we compare the performances of the considered scheme with those of
CP-OFDMA in a simple study case: the channels of both users do not introduce
other impairment than a delay. More specifically, we consider the receiver to be
perfectly synchronized with User 1, while User 2 is received with a delay equal
to an integer number of samples. In what follows, we focus on how the delay of
User 2 affects the SNDR of User 1. Note that, analogously, we could have focused
on User 2. Indeed, even though the two users cannot synchronize to one another,
it is realistic to assume that the BS is able to synchronize alternatively to
both of them in order to retrieve their transmitted signal. The ability of FBMC
to deal with the timing of each subcarrier (and hence of each user) separately
is a strong advantage of this multicarrier multiple-access scheme when compared
to classical CP-OFDM.

\begin{figure}
\centering
\begin{tabular}{rr}
  \includegraphics{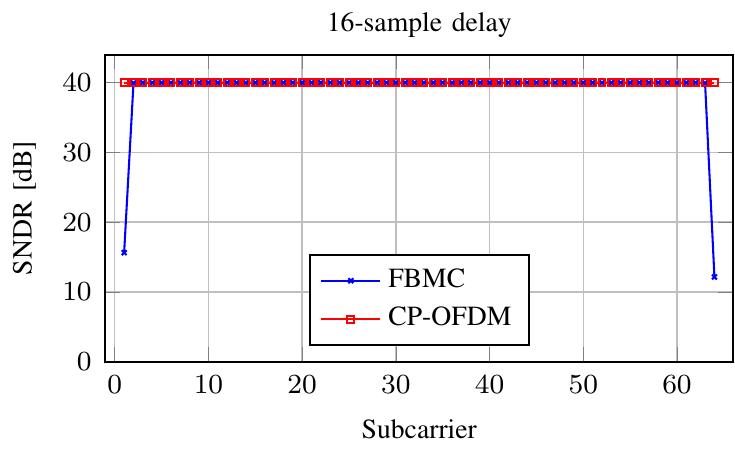}&
  \includegraphics{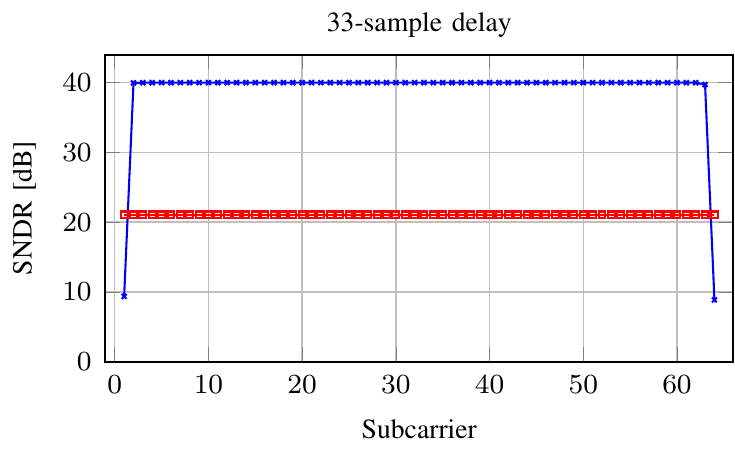}\\
  \includegraphics{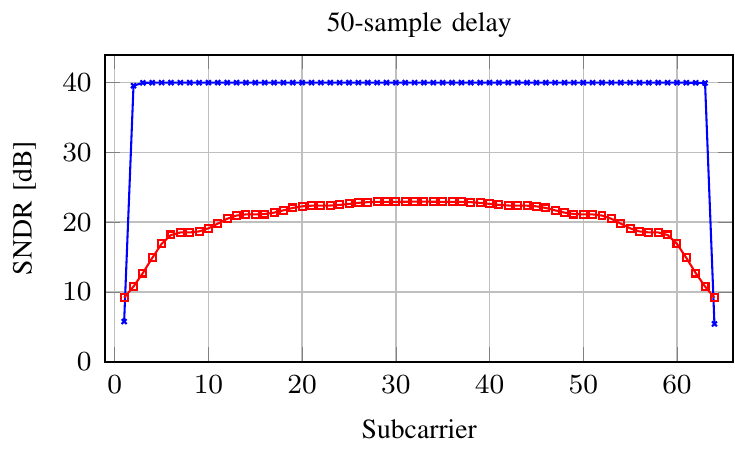}&
  \includegraphics{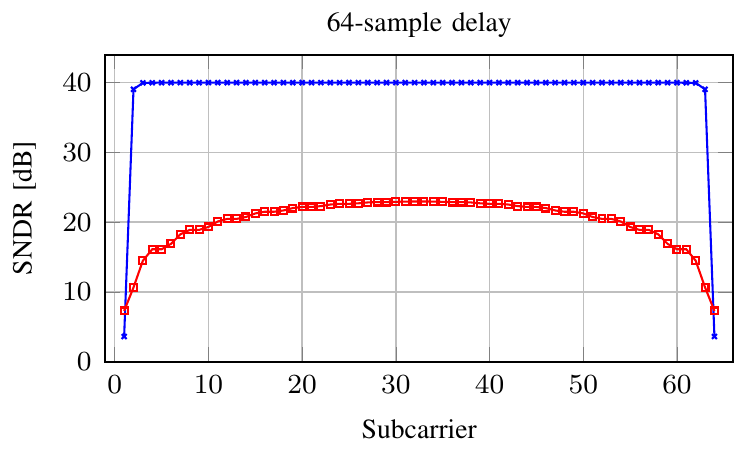}
\end{tabular}

\caption{Comparison between FBMC-based OFDMA and CP-OFDMA: SNDR of User 1 for
different delay values of User 2. The transmit SNR is set to 40~dB.}\label{fig:vsOFDMA}
\end{figure}

Fig.~\ref{fig:vsOFDMA} shows the SNDR corresponding to the subcarriers of User 1
for different delays between the two users. The transmission scheme is either
the FBMC/OQAM-based one described in the previous sections or a classic CP-OFDMA
scheme with a cyclic prefix of 32 samples (25\% of the number of subcarriers).
Both users are assumed to transmit with a power that is 40~dB above the noise
level. When the delay is shorter than the CP, CP-OFDMA is capable of
compensating it perfectly and no distortion is introduced (see
Fig.~\ref{fig:vsOFDMA}a corresponding to a delay of 16 samples). However, as
soon as the delay between the two users is larger than the CP (see
graphs b, c and d of Fig.~\ref{fig:vsOFDMA}) the performances of CP-OFDMA decay
catastrophically, with a drop of 20~dB. A detailed analysis of the distortion
generated in a CP-OFDMA system is reported in Appendix~\ref{sec:dist_OFDM}.

Conversely, FBMC-based OFDMA proves itself robust to user asynchronicities: User
1 is received with the same SNDR in all the considered examples. More
specifically, for all delay values, the quality of the received signal is
approximately the same as the one obtained by CP-OFDMA in the short delay case
(see Fig.~\ref{fig:vsOFDMA}a). Only the first and the last subcarriers show a
significant degradation, due to the leakage effect discussed before.

\subsection{Symbol Error Rate}
\begin{figure}
  \centering
  \begin{tabular}{rr}
  \hspace{-3mm}%
  \includegraphics{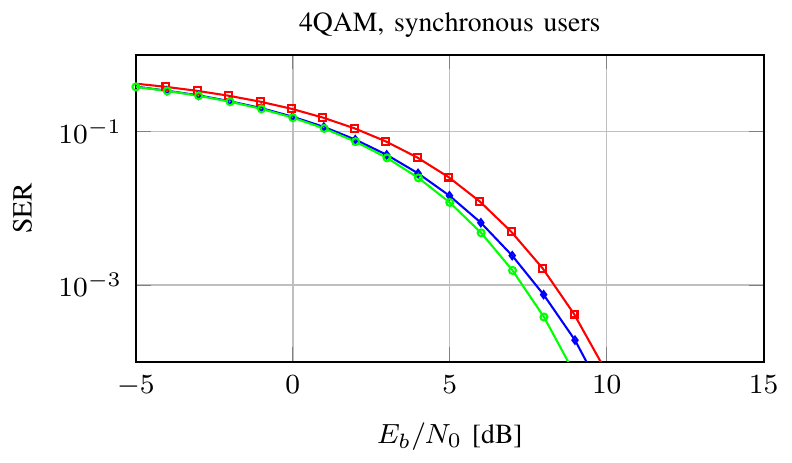}
  &\hspace{-5mm}
  \includegraphics{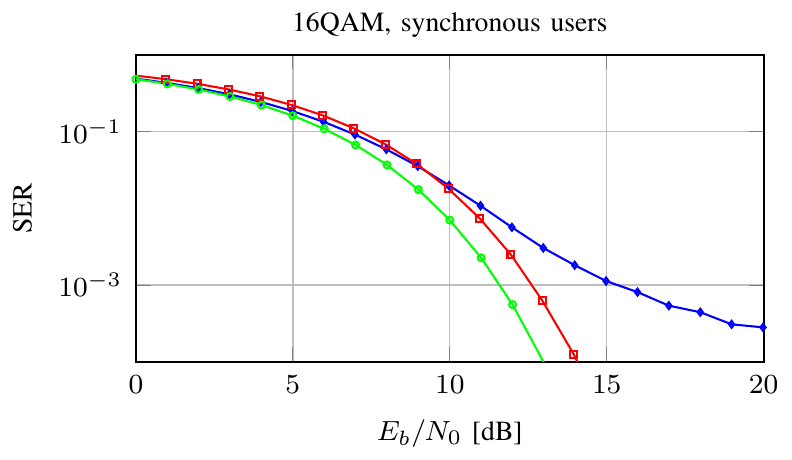}\\
  \hspace{-3mm}
  \includegraphics{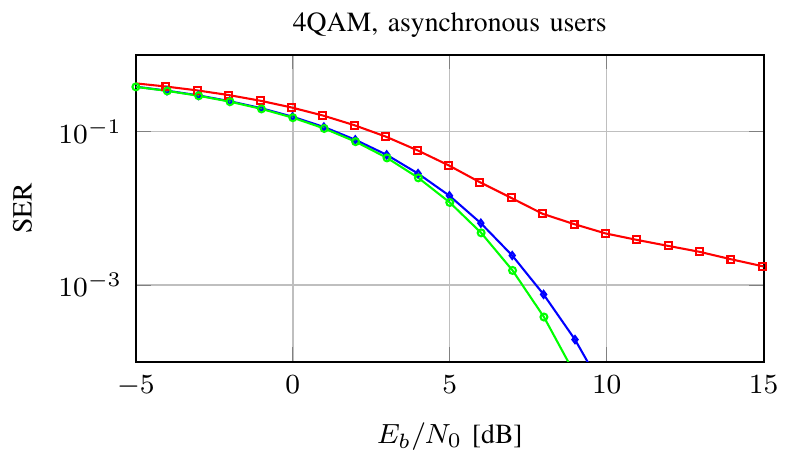}
  &\hspace{-5mm}
  \includegraphics{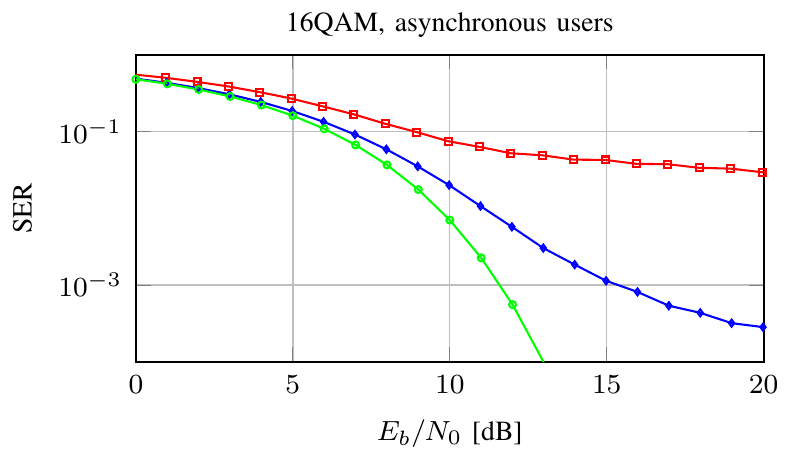}\\
  \multicolumn{2}{c}{\includegraphics{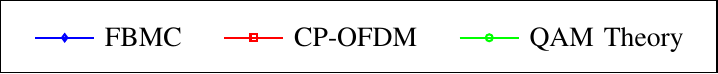}}
  \end{tabular}
  \caption{Symbol error rate as a function of $E_b/N_0$ (measured at the
    receiver side) for FBMC-based OFDMA and
  CP-OFDMA. Users' transmissions are either synchronized (top) or not (bottom)
and the transmitted symbols belong either to a 4QAM constellation (left) or to a
16QAM constellation (right).}\label{fig:SER}
\end{figure}

To conclude our study and the comparison between the two multiple-access
techniques, Fig.~\ref{fig:SER} reports some Symbol-Error-Rate (SER) curves obtained
simulating the transmission of 1000 bursts of 100 multicarrier symbols. For each
burst, the users' channels are drawn independently according to the EVA model.
SER is reported as a function of the received energy per bit $E_b$, normalized
with respect to the noise spectral density~$N_0$. A single guard band separates
the users (subcarriers 64 and 128 are switched off).

Two different synchronization assumptions are considered. In the first one (top
graphs), the transmissions of the two users are synchronized. In this case,
CP-OFDMA follows the theoretic performance of the chosen QAM constellation (4QAM
for the left graphs and 16QAM for the right ones \cite[Chapter
5]{Benedetto_Principles}) with a 1-dB loss, approximately, due to the CP
overhead. Indeed, the length of the channel (around 14 taps for the EVA
model with a 1.92-MHz bandwidth) is perfectly compensated by the 32-sample CP.
Conversely, FBMC is closer to the theoretic curve at low $E_b/N_0$ values, but
its performance worsens as we increase the SNR due to the inherent interference
discussed above, as suggested by the curve floor visible in the 16QAM case.
(In the 4QAM case, the SER floor of the FBMC curve falls
outside the depicted range.)

The bottom graphs, on the other hand, are obtained introducing a delay between
the two users. This delay is equal to an integer number of samples, varies at
each burst and is uniformly distributed between 0 and 127 samples (i.e.\ one
multicarrier symbol). As we can see from Fig.~\ref{fig:SER}, asynchronous
transmissions do not imply a significant performance loss for the FBMC configuration.
This is not the case for the scheme based on CP-OFDM: since the CP is too short
to compensate the total length of the channel (delay plus impulse response), the
SER does not improve as desired for $E_b/N_0>10$~dB and the performance is much
worse than for the FBMC scheme.

It is worth remarking that the FBMC curves are obtained with a suboptimal
one-tap per-subcarrier equalizer. The performance gap between FBMC-based OFDMA
and CP-OFDMA will increase by employing receiver structures such as
\cite{Ihalainen2007, Mestre2013}, which will lower the FBMC error floor. Even if
the equalizers described in \cite{Ihalainen2007, Mestre2013} are more complex
than a CP-OFDM one-tap per-subcarrier equalizer, they are still a competitive
solution when compared with hardware and protocol requirements of classic OFDMA
with short CP \cite{Morelli_etal}. 

\section{Conclusions}\label{sec:conclusions}
In this paper we have rigorously analyzed the interaction between users in an
FBMC/OQAM-based OFDMA system. More specifically, for a reasonably high number of
subcarriers, we have been able to express the per-subcarrier interference as the
sum of users' terms. It is important to remark that, in general, the distortion
at a given subcarrier does not depend exclusively on the user transmitting on
that subcarrier, but also on all other users. Our analysis shows that, when the prototype
pulses are well designed, the distortion caused by a given user decays steeply
outside the user's subband and can be neglected after few subcarriers (or even
after a single subcarrier, as in the example of Fig.~\ref{fig:distortion}). This
result supports the choice of placing a single empty guard band between users.
To the best of our knowledge, such a choice was only empirically justified until
now. Moreover, the spread of the out-band interference depends neither on the
channel responses (of any user) nor on the equalizer coefficients, thus
simplifying the task of designing the prototype pulses.

As for the comparison to the more classical CP-OFDMA scheme, the ideal case
with synchronized users suggests that the FBMC approach is not
worth the complexity. Indeed, the CP causes a loss in the order of 1~dB at low-to-medium
SNR, but avoids performance floors at high SNR (see Fig.~\ref{fig:SER}, top
graphs). However, in more practical scenarios where users' transmissions are not
synchronized, the length of the equivalent channel (delay plus impulse response)
can easily exceed the CP. In this situation, the performance of CP-OFDMA drops
abruptly, whereas the FBMC scheme proves to be very robust, showing no
significant degradation and without need for any extra signal processing
(see Fig.~\ref{fig:SER}, bottom graphs).

Summarizing, filterbank modulations offer a convenient solution to the
multiple-access channel problem, since they can deal with asynchronous users
without increasing complexity. Furthermore, the tools presented in this paper
allow the designer to choose the prototype pulses that minimize both inter-user
interference and number of guard bands. Remarkably, no channel state information
is needed to carry out this task.

\appendices
\section{Proof of Proposition~\ref{prop:dist_terms}}\label{sec:proof_dist}
In this appendix we show how the terms in (\ref{eq:dist_terms}) are derived. To
do so, we first obtain a general expression for the distortion generated by user
$k$ at subcarrier $m$. Then, the simplified form in (\ref{eq:dist_terms}) is
obtained by particularizing to the the special case considered in
Proposition~\ref{prop:dist_terms}.

\subsection{The General Case}
For practical reasons, let us define the following matrices and vectors:
\begin{align}
  \bm y_k^{\odd}\Bigl(p_N,q_N^{(r)}\Bigr) &= {\Bigl[\bm
  Y_k^\odd\Bigl(p_N,q_N^{(r)}\Bigr)\Bigr]}_{:,n+\kappa-1}\nonumber\\
  \overline{\bm y_k^{\odd}}\Bigl(p_N,q_N^{(r)}\Bigr) &= \begin{bmatrix}
  \Re\Bigl\{\bm y_k^{\odd}\Bigl(p_N,q_N^{(r)}\Bigr)\Bigr\}\\
  -\Im\Bigl\{\bm y_k^{\odd}\Bigl(p_N,q_N^{(r)}\Bigr)\Bigr\}\end{bmatrix}\nonumber\\
  t_{k,m}(i) &= \frac{(-\cj)^i}{i!} W_m H_k^{(i)}(\omega_m),\hspace{2em}i=0,1,2 \nonumber\\
  \bm T_k(i) &= \diag_{m=1,\dots,2M}{\{t_{k,m}(i)\}} \nonumber\\
   &= \frac{(-\cj)^i}{i!}\bm W \bs
  \Lambda_{H_k^{(i)}},\hspace{2em}i=0,1,2\nonumber\\
  \bm t_{k,m}(i) &= \begin{bmatrix} \bm 0_{m-1} \\ \Re\{t_{k,m}(i)\}\\ \bm
  0_{2M} \\ \Im\{t_{k,m}(i)\} \\ \bm 0_{2M-m-1}
  \end{bmatrix}\label{eq:def_t} \\
  \intertext{and, for a generic column vector $\bm u\in\mathbb{C}^{2M}$,}
  \mathcal{M}(\bm u) &= \bm\Phi \bm F_{2M}^H \diag\bigl\{\bm F_{2M} \bm\Phi^*\bm
  u\bigr\} \nonumber\\
  \widetilde{\mathcal{M}}(\bm u) &= \bm\Phi \bm F_{2M}^H \diag\bigl\{(\bm J_2
  \otimes \bm I_M)\bm F_{2M} \bm\Phi^*\bm u\bigr\}.\nonumber
\end{align}
Then, denoting by $\bm b_k(r)$ [$\bm c_k(r)$, respectively] the $r$-th column of
$\bm B_k$ ($\bm C_k$, respectively), from (\ref{eq:ideal_comp}) one can write 
\begin{multline}\label{eq:y_odd}
\bm y_k^{\odd}\Bigl(p_N,q_N^{(r)}\Bigr) = 2\sum_{i=1}^{N_s} \mathcal{M}
\bigl(\bm b_k(i)\bigr) {[\mathcal{R}(p_N,q_N^{(r)})]}_{:,n+\kappa-i}\\
{}+2\sum_{i=1}^{N_s}\widetilde{\mathcal{M}}\bigl(\cj \bm c_k(i)\bigr)
  \begin{bmatrix}
    {[\mathcal{S}_1(p_N,q_N^{(r)}]}_{:,n+\kappa-1-i}\\
    {[\mathcal{S}_2(p_N,q_N^{(r)}]}_{:,n+\kappa-i}
  \end{bmatrix}
\end{multline}
where all column vectors are intended identically null when the column index is negative or
higher than $2\kappa-1$. Moreover, keeping (\ref{eq:freqsel}) and (\ref{eq:def_t}) in
mind, the distortion can be expressed as follows:
\begin{multline*}
  \EV\biggl[{\Bigl|\Re[z_{k,m}^\odd]- t_{k,m}(0) b_{k,m}\Bigr|}^2\biggr] = 
\eta_{0,0}(m) + \frac{2}{2M}\eta_{0,1}(m) \\
{}+\frac{1}{4M^2}\eta_{0,2}(m) + \frac{1}{4M^2}\eta_{1,1}(m) +
o\bigl(M^{-2}\bigr)
\end{multline*}
where
\begin{align*}
  \eta_{0,0}(m) = \:& \bm t_{k,m}^T(0) \EV \biggl[
\Bigl(\overline{\bm y^{\odd}_k}(p_N,q_N) - {[\bm b_k^T(n), \bm 0_{2M}^T]}^T
\Bigr)
\Bigl(\overline{\bm y^{\odd}_k}(p_N,q_N) - {[\bm b_k^T(n), \bm 0_{2M}^T]}^T
\Bigr)^T
\biggr] \bm t_{k,m}(0)\\
\eta_{0,1}(m) =\:& \bm t_{k,m}^T(0) \EV \biggl[
\Bigl(\overline{\bm y^{\odd}_k}(p_N,q_N) - {[\bm b_k^T(n), \bm 0_{2M}^T]}^T
\Bigr)
\times\Bigl(\overline{\bm y^{\odd}_k}(p_N,q_N') \Bigr)^T \biggr] \bm t_{k,m}(1)\\
\eta_{0,2}(m) =\:& \bm t_{k,m}^T(0) \EV \biggl[
\Bigl(\overline{\bm y^{\odd}_k}(p_N,q_N) - {[\bm b_k^T(n), \bm 0_{2M}^T]}^T
\Bigr)
\Bigl(\overline{\bm y^{\odd}_k}(p_N,q_N'') \Bigr)^T \biggr] \bm t_{k,m}(2)\\
\eta_{1,1}(m) =\:& \bm t_{k,m}^T(1) \EV \biggl[
\Bigl(\overline{\bm y^{\odd}_k}(p_N,q_N')\Bigr)
\Bigl(\overline{\bm y^{\odd}_k}(p_N,q_N') \Bigr)^T \biggr] \bm t_{k,m}(1).
\end{align*}

Before taking the next step, we need to introduce the following result, whose
proof follows the same lines as \cite[Lemma 2]{Mestre2013}.

\begin{figure*}
\begin{subequations}\label{eq:eta_long}
\begin{align}
  \eta_{0,0}(m) =\:&
    2 \Re^2[t_{k,m}(0)] \Re \tr\Bigl[\bm U^+ \mathcal F_k(m)
      \mathcal{X}^{\mathcal R-\frac{1}{2}\mathbb I}_{(p,q,p,q)} + \bm U^- \mathcal F_k(m) \mathcal{X}^{\mathcal
  S}_{(p,q,p,q)}\Bigr]\nonumber\\
  &{}+2\Im^2[t_{k,m}(0)] \Re\tr\Bigl[\bm U^- \mathcal F_k(m)
\mathcal{X}^{\mathcal R-\frac{1}{2}\mathbb I}_{(p,q,p,q)} + \bm U^+ \mathcal F_k(m) \mathcal{X}^{\mathcal
S}_{(p,q,p,q)}\Bigr]\nonumber\\
&{} -4\Re[t_{k,m}(0)]\Im[t_{k,m}(0)]\nonumber\\
&\qquad{} \times\Im\tr\Bigl[(\bm I_2 \otimes \bm J_M) \mathcal F_k(m)
\mathcal{X}^{\mathcal R-\frac{1}{2}\mathbb I}_{(p,q,p,q)} -(\bm I_2 \otimes \bm J_M) \mathcal F_k(m) \mathcal{X}^{\mathcal
S}_{(p,q,p,q)}\Bigr]\label{eq:eta00long}\\
  \eta_{0,1}(m) =\:&
    2 \Re[t_{k,m}(0)]\Re[t_{k,m}(1)] \Re \tr\Bigl[\bm U^+ \mathcal F_k(m)
      \mathcal{X}^{\mathcal R-\frac{1}{2}\mathbb I,\mathcal R}_{(p,q,p,q')} + \bm U^- \mathcal F_k(m) \mathcal{X}^{\mathcal
  S}_{(p,q,p,q')}\Bigr]\nonumber\\
  &{}+2\Im[t_{k,m}(0)]\Im[t_{k,m}(1)] \Re\tr\Bigl[\bm U^- \mathcal F_k(m)
\mathcal{X}^{\mathcal R-\frac{1}{2}\mathbb I,\mathcal R}_{(p,q,p,q')} + \bm U^+ \mathcal F_k(m) \mathcal{X}^{\mathcal
S}_{(p,q,p,q')}\Bigr]\nonumber\\
&{}+2\Re[t_{k,m}(0)]\Im[t_{k,m}(1)] \Im\tr\Bigl[\bm U^- \mathcal F_k(m)
\mathcal{X}^{\mathcal R-\frac{1}{2}\mathbb I,\mathcal R}_{(p,q,p,q')} +\bm U^+ \mathcal F_k(m) \mathcal{X}^{\mathcal
S}_{(p,q,p,q')}\Bigr]\nonumber\\
&{}-2\Im[t_{k,m}(0)]\Re[t_{k,m}(1)] \Im\tr\Bigl[\bm U^+ \mathcal F_k(m)
\mathcal{X}^{\mathcal R-\frac{1}{2}\mathbb I,\mathcal R}_{(p,q,p,q')} +\bm U^- \mathcal F_k(m) \mathcal{X}^{\mathcal
S}_{(p,q,p,q')}\Bigr]\label{eq:eta01long}\\
  \eta_{0,2}(m) =\:&
    2 \Re[t_{k,m}(0)]\Re[t_{k,m}(2)] \Re \tr\Bigl[\bm U^+ \mathcal F_k(m)
      \mathcal{X}^{\mathcal R-\frac{1}{2}\mathbb I,\mathcal R}_{(p,q,p,q'')} + \bm U^- \mathcal F_k(m) \mathcal{X}^{\mathcal
  S}_{(p,q,p,q'')}\Bigr]\nonumber\\
  &{}+2\Im[t_{k,m}(0)]\Im[t_{k,m}(2)] \Re\tr\Bigl[\bm U^- \mathcal F_k(m)
\mathcal{X}^{\mathcal R-\frac{1}{2}\mathbb I,\mathcal R}_{(p,q,p,q'')} + \bm U^+ \mathcal F_k(m) \mathcal{X}^{\mathcal
S}_{(p,q,p,q'')}\Bigr]\nonumber\\
&{}+2\Re[t_{k,m}(0)]\Im[t_{k,m}(2)] \Im\tr\Bigl[\bm U^- \mathcal F_k(m)
\mathcal{X}^{\mathcal R-\frac{1}{2}\mathbb I,\mathcal R}_{(p,q,p,q'')} +\bm U^+ \mathcal F_k(m) \mathcal{X}^{\mathcal
S}_{(p,q,p,q'')}\Bigr]\nonumber\\
&{}-2\Im[t_{k,m}(0)]\Re[t_{k,m}(2)] \Im\tr\Bigl[\bm U^+ \mathcal F_k(m)
\mathcal{X}^{\mathcal R-\frac{1}{2}\mathbb I,\mathcal R}_{(p,q,p,q'')} +\bm U^- \mathcal F_k(m) \mathcal{X}^{\mathcal
S}_{(p,q,p,q'')}\Bigr]\label{eq:eta02long}\\
\eta_{1,1}(m) =\:&
    2 \Re^2[t_{k,m}(1)] \Re \tr\Bigl[\bm U^+ \mathcal F_k(m)
    \mathcal{X}^{\mathcal R}_{(p,q',p,q')} + \bm U^- \mathcal F_k(m) \mathcal{X}^{\mathcal
  S}_{(p,q',p,q')}\Bigr]\nonumber\\
  &{}+2\Im^2[t_{k,m}(1)] \Re\tr\Bigl[\bm U^- \mathcal F_k(m)
\mathcal{X}^{\mathcal R}_{(p,q',p,q')} + \bm U^+ \mathcal F_k(m) \mathcal{X}^{\mathcal
S}_{(p,q',p,q')}\Bigr]\nonumber\\
&{}-4\Re[t_{k,m}(1)]\Im[t_{k,m}(1)]\nonumber\\&\qquad{}\times \Im\tr\Bigl[(\bm I_2 \otimes \bm J_M) \mathcal F_k(m)
\mathcal{X}^{\mathcal R}_{(p,q',p,q')} -(\bm I_2 \otimes \bm J_M) \mathcal F_k(m) \mathcal{X}^{\mathcal
S}_{(p,q',p,q')}\Bigr]\label{eq:eta11long}
\end{align}
\end{subequations}

\hfill\rule{.99\textwidth}{.5pt}\hfill\null
\end{figure*}

\begin{lemma} \label{lemma:M_prods}
  Let $\bm u\in\mathbb{R}^{2M}$ be a real-valued random vector of independent
  entries with zero mean and diagonal covariance matrix $\bm Q_u$. Recalling the
  definitions of $\bm U^+$ and $\bm U^-$ in (\ref{eq:Udef}), and for $m=1,\dots,2M$, we can write
  \begin{align*}
    \EV\biggl[{\bigl[\mathcal{M}^{\Re}(\bm u)\bigr]}_{m,:}^T
    {\bigl[\mathcal{M}^{\Re}(\bm u)\bigr]}_{m,:}\biggr] &= \frac{1}{2}\bm U^+
      \Re\{\mathcal{F}_u(m)\}\\
    \EV\biggl[{\bigl[\mathcal{M}^{\Im}(\bm u)\bigr]}_{m,:}^T
    {\bigl[\mathcal{M}^{\Im}(\bm u)\bigr]}_{m,:}\biggr] &= \frac{1}{2}\bm U^-
      \Re\{\mathcal{F}_u(m)\}\\
    \EV\biggl[{\bigl[\mathcal{M}^{\Re}(\bm u)\bigr]}_{m,:}^T
    {\bigl[\mathcal{M}^{\Im}(\bm u)\bigr]}_{m,:}\biggr] &= \frac{1}{2}\bm U^+
      \Im\{\mathcal{F}_u(m)\}\\
    \EV\biggl[{\bigl[\mathcal{M}^{\Im}(\bm u)\bigr]}_{m,:}^T
    {\bigl[\mathcal{M}^{\Re}(\bm u)\bigr]}_{m,:}\biggr] &= -\frac{1}{2}\bm U^-
      \Im\{\mathcal{F}_u(m)\}
  \end{align*}
  where $\mathcal{F}_u(m) = \bm F_{2M}^H \bm Q_u \bm F_{2M} \odot \bm f(m) \bm
  f^H(m)$, together with $\mathcal{M}^{\Re}(\bm u)=\Re\{\mathcal{M}(\bm u)\}$ and
  $\mathcal{M}^{\Im}(\bm u)=\Im\{\mathcal{M}(\bm u)\}$.

  The same results hold when $\mathcal{M}(\bm u)$ is replaced everywhere by
  $\widetilde{\mathcal M}(\bm u)$.
\end{lemma}

Given (\ref{eq:y_odd}), a direct application of this lemma to $\eta_{1,1}(m)$ yields
(\ref{eq:eta11long})
at the top of the page. Recall that $\mathcal{X}^{\mathcal R}_{(p,q',p,q')}$ and $\mathcal{X}^{\mathcal
S}_{(p,q',p,q')}$ are defined in (\ref{eq:Wr}) and (\ref{eq:Xs}), respectively,
and that $\mathcal{F}_k(m) = \bm F_{2M}^H \bm Q_k \bm F_{2M} \odot \bm f(m) \bm
f^H(m)$, with $\bm Q_k = \EV[\bm b_k(r) \bm b_k^T(r)] = \EV[\bm c_k(r) \bm
c_k^T(r)]$, for any $r=1,\dots,N_s$.

Similar expressions can be found for the other three terms $\eta_{0,0}(m)$,
$\eta_{0,1}(m)$ and $\eta_{0,2}(m)$, also [see
(\ref{eq:eta00long})--(\ref{eq:eta02long})]. Indeed, it is enough to realize that
(\ref{eq:y_odd}) holds true even if we replace $\bm y_k^{\odd}(p_N,q_N)$ by $\bm
y_k^{\odd}(p_N,q_N) - \bm b_k(n)$ and $\mathcal R(p_N,q_N)$ by $\mathcal
R(p_N,q_N) - \frac{1}{2}\mathbb I$.

\subsection{The Special Case of Proposition \ref{prop:dist_terms}}
In the previous section we have derived a general expression for the distortion
caused by user $k$ at subcarrier $m$ and shown how it depends on
\begin{inparaenum}
\item the power/resource allocation policy through matrix $\mathcal{F}_k(m)$,
\item the channel and the chosen equalizer through coefficients $t_{k,m}(i)$, and 
\item the prototype pulses through matrices $\mathcal R(p_N,q_N^{(r)})$ and
  $\mathcal S(p_N,q_N^{(r)})$.
\end{inparaenum}
It is worth remarking that the expression is as general as possible and holds
for any choice of prototype pulses $p_N[n]$ and $q_N[n]$ that 
fulfill assumptions \textbf{\textsl{AS1}} and \textbf{\textsl{AS2}}. More specifically, there are no 
requirements about symmetry or perfect reconstruction. In what follows, we show how the
distortion terms simplify to~(\ref{eq:dist_terms}) when the prototype pulses are symmetric
(e.g.\ $p_N[n]=p_N[N-n+1]$) and meet PR conditions (\ref{eq:RC}).

To that purpose, we need the two following results. Proofs are omitted to due to
space constraints, but are straightforward consequences of the fact that
$\mathcal F_k(m)$ is a circulant matrix.

\begin{lemma}\label{lemma:circ1}
  For any $2M\times 2M$ complex matrix $\bm A$ the following identities hold
  true (note the commuting signs in the second equation):
\begin{align*}
\Re \tr\Bigl[\bm U^\pm \mathcal{F}_k(m) \bm A\Bigr] &= \frac{1}{2} \Re
\tr\Bigl[\mathcal{F}_k(m) \bm U^\pm \bm A \bm U^\pm \Bigr] \\
\Im \tr\Bigl[\bm U^\pm \mathcal{F}_k(m) \bm A\Bigr] &= \frac{1}{2} \Im
\tr\Bigl[\mathcal{F}_k(m) \bm U^\mp \bm A \bm U^\pm \Bigr].
\end{align*}
\end{lemma}

\begin{lemma}\label{lemma:circ2}
  Let $\bm A$ be a $2M\times 2M$ complex matrix such that
  $$
  \bm A = \begin{bmatrix} \bm A_1 & \bm A_2 \\
    -\bm A_2 & -\bm A_1 \end{bmatrix}.
  $$
  Then
  $$
  \tr\Bigl[\mathcal F_k(m) \bm A\Bigr] = 0.
  $$
\end{lemma}

Lemma~\ref{lemma:circ1}, together with (\ref{eq:RC}), implies that most of the
terms in (\ref{eq:eta_long}) cancel out when the prototype pulses meet the PR
conditions. Take, for instance, the first term of $\eta_{0,0}(m)$. We have
$$
2\Re \tr\Bigl[\bm U^+ \mathcal F_k(m) \mathcal{X}^{\mathcal R-\frac{1}{2}\mathbb
I}_{(p,q,p,q)}\Bigr] = \Re \tr\Bigl[\mathcal F_k(m) \bm U^+ \mathcal{X}^{\mathcal
R-\frac{1}{2}\mathbb I}_{(p,q,p,q)} \bm U^+\Bigr]=0
$$
where we have used the fact that (\ref{eq:RC1}) can be rewritten as $\bm U^+
\Bigl(\mathcal{R}(p_N,q_N) - \frac{1}{2}\mathbb{I}\Bigr) = \bm 0$.

Assume now that the prototype pulses are also symmetric, besides PR-compliant.
Note that if $q_N[n]$ is symmetric (i.e.\ $q_N[n]=q_N[N-n+1]$), then so is its
second derivative (i.e.\ $q_N''[n] = q_N''[N-n+1]$), while the first derivative
is anti-symmetric (i.e.\ $q_N'[n] = -q_N'[N-n+1]$). In this case, other terms
null out as a consequence of Lemma~\ref{lemma:circ2}. For example, for the
second term of $\eta_{0,1}(m)$, we have
$$
  2\Re\tr\Bigl[\bm U^- \mathcal F_k(m) \mathcal{X}^{\mathcal R-\frac{1}{2}\mathbb
  I,\mathcal R}_{(p,q,p,q')} \Bigr] = 
  \Re\tr\Bigl[\mathcal F_k(m) \bm U^- \mathcal{X}^{\mathcal R-\frac{1}{2}\mathbb
  I,\mathcal R}_{(p,q,p,q')} \bm U^- \Bigr] = \bm 0
$$
since the symmetry properties of $p_N[n]$ and $q_N[n]$, together with $\bm U^-
\mathbb I = 0$, imply that product matrix $\bm U^- \mathcal{X}^{\mathcal R-\frac{1}{2}\mathbb
I,\mathcal R}_{(p,q,p,q')} \bm U^-$ meets the hypothesis of
Lemma~\ref{lemma:circ2}.

We have thus proven how the general distortion terms in (\ref{eq:eta_long})
simplify into those of (\ref{eq:dist_terms}) when the prototype pulses are
symmetric and satisfy the PR conditions.

\subsection{The Single-User Case}
The distortion formula (\ref{eq:dist_formula}) with terms $\eta_{x,y}(m)$ given
by~(\ref{eq:eta_long}) holds for all prototype pulses and for any number of
users. In particular, for $K=1$, it allows computing the distortion of a
single-user FBMC/OQAM link with a one-tap equalizer per subcarrier. The
resulting expression
\begin{multline}\label{eq:1user}
P_e(m) =
\frac{P_s}{M}\tr\Bigl[\bm U^+ \mathcal{X}^{\mathcal R - \frac{1}{2}
\mathbb{I}}_{(p,q,p,q)}\Bigr]
+\frac{P_s}{M^2}\Im\Bigl[W_m H^{(1)}(\omega_m)\Bigr]\tr \Bigl[\bm
U^+\mathcal{X}^{\mathcal R - \frac{1}{2}
\mathbb{I},\mathcal{R}}_{(p,q,p,q')}\Bigr]\\
{}-\frac{P_s}{4M^3}\Re\Bigl[W_m H^{(2)}(\omega_m)\Bigr]\tr\Bigl[\bm
U^+\mathcal{X}^{\mathcal R - \frac{1}{2} \mathbb{I},
\mathcal{R}}_{(p,q,p,q")}\Bigr]\\
{}+\frac{P_s}{4M^3}{\Bigl|W_m H^{(1)}(\omega_m)\Bigr|}^2\tr\biggl[ \bm
  U^+\mathcal{X}^{\mathcal{R}}_{(p,q',p,q')} + \bm U^-\mathcal{X}^{\mathcal
S}_{(p,q',p,q')}\biggr]
\end{multline}
is thus a generalization of \cite[Eq.
(35)]{Mestre2013}, which refers to the PR case.

\section{Distortion in CP-OFDM}\label{sec:dist_OFDM}
This appendix derives an expression for the signal distortion in a CP-OFDMA
system where the length of the channel is larger than the cyclic prefix. More
specifically, we consider a CP of length $L$ and a channel of length $L+L_T$ with taps
$\{h_{k,\ell}\}_{\ell=1,\dots,L+L_T}$ (for simplicity, we assume here that
$L_T<L<2M$). The resulting distortion
expression was used to draw the CP-OFDM curves in Fig.~\ref{fig:vsOFDMA}.

Maintaining the same notation as in Section~\ref{sec:signal_model}, the
contribution to the output signal due to user $k$ can be written as
\begin{equation}\label{eq:OFDM_model}
\bm Z_k^{\text{(CP)}} = \bm W\bm F_{2M}^H \overline{\bm \Psi} \bm H_k \bm \Psi \bm F_{2M}
[ \bm A_k \; \bm 0] + \bm W\bm F_{2M}^H \overline{\bm \Psi} \bm
H_{k,T} \bm \Psi \bm F_{2M} [ \bm 0 \; \bm A_k]
\end{equation}
where matrices
\begin{align*}
  \bm \Psi &= \begin{bmatrix}\bm 0_{L\times(2M-L)}\;\bm I_L\\\bm
I_{2M}\end{bmatrix} &
\overline{\bm \Psi} &= \begin{bmatrix}\bm 0_{2M\times L}\; \bm I_{2M}\end{bmatrix}
\end{align*}
apply and remove the CP of length $L$, respectively. The $(2M+L)\times(2M+L)$
channel matrix $\bm H_k$ is a lower triangular Toeplitz matrix whose first column
is $[h_{k,1}\;\cdots\;h_{k,L+L_T}\;\bm 0_{1\times(2M-L_T)}]^T$. Similarly, the
$(2M+L)\times(2M+L)$ matrix $\bm H_{k,T}$ is a upper triangular Toeplitz matrix
whose first row is $[\bm 0_{1\times(2M-L_T+1)}\;h_{k,L+L_T}\;\cdots\;h_{k,2}]$.

It is straightforward to realize that (\ref{eq:OFDM_model}) corresponds to the
classic OFDM model when $L_T=1$ and the CP covers the entire channel length.
Indeed, $L_T=1$ implies $\bm H_{k,T}=\bm 0$. Also, we have
$$
\bm\Lambda_{H_k} = \bm F_{2M}^H \overline{\bm \Psi} \bm H_k \bm \Psi \bm F_{2M}
$$
where $\bm\Lambda_{H_k}$ is the diagonal matrix filled with the channel
frequency response defined in Section~\ref{sec:signal_model}. Thus, for channels
not longer than $L+1$ taps, it is enough to set $\bm W=\bm\Lambda_{H_k}^{-1}$ to
recover the transmitted symbols perfectly, namely $\bm Z_k^{\text{(CP)}}=[\bm
A_k\;\bm 0]$.

Conversely, when $L_T>1$, matrix $\overline{\bm \Psi} \bm H_k \bm \Psi$ is not
diagonalizable anymore and an extra term appears. Namely
$$
\overline{\bm \Psi} \bm H_k \bm \Psi = \bm F_{2M} \bm \Lambda_{H_k} \bm F_{2M}^H
-\widetilde{\bm H}_k
$$
where we introduced the $2M\times 2M$ matrix
\begin{equation}\label{eq:Hk}
\widetilde{\bm H}_k = \begin{bmatrix}\bm 0_{(L_T-1)\times (2M-L-L_T+1)} & \bm T_k & \bm 0_{(L_T-1)\times L}\\&
\makebox[1.5em][l]{\hspace*{-2em}$\bm0_{(2M-L_T+1)\times 2M}$}\end{bmatrix}
\end{equation}
and where $\bm T_k$ is a $(L_T-1)\times(L_T-1)$ upper triangular Toeplitz matrix
with $[h_{k,L+L_T}\;\cdots\;h_{k,L+2}]$ as its first row. Furthermore, let us denote
\begin{equation}\label{eq:HkT}
\widetilde{\bm H}_{k,T} = \overline{\bm \Psi} \bm H_{k,T} \bm \Psi = 
\begin{bmatrix}\bm 0_{(L_T-1)\times (2M-L_T+1)} & \bm T_k\\
\makebox[1.5em][l]{\hspace*{-1em}$\bm0_{(2M-L_T+1)\times 2M}$}\end{bmatrix}.
\end{equation}
Then, for a generic one-tap-per-subcarrier equalizer $\bm W$, (\ref{eq:OFDM_model}) can be rewritten as
\begin{equation}\label{eq:CP_signal_dist}
\bm Z_k^{\text{(CP)}} = \bm W \bm \Lambda_{H_k} [\bm A_k\;\bm 0] - \bm W\bm
F_{2M}^H \widetilde{\bm H}_k \bm F_{2M} [ \bm A_k \; \bm 0]
\bm W\bm F_{2M}^H
\widetilde{\bm H}_{k,T} \bm F_{2M} [ \bm 0 \; \bm A_k].
\end{equation}

\begin{proposition}
  Let $\bm a_k$ be a generic column of matrix $\bm A_k$ and let $\bs \sigma_k =
  (2M)^{-\frac{1}{2}}\bm F^H \diag\EV[\bm a_k\bm a_k^H]$ (independent of the choice of $\bm a_k$
  since all entries of $\bm A_k$ are i.i.d.). In other words, the elements
  $\{\sigma_{k,i}\}_{i=1,\dots,2M}$ of $\bs \sigma_k$ are the normalized Fourier coefficients
  associated to the power levels assigned to the subcarriers by user $k$. Next,
  denote by $\bm X_k$ the Toeplitz upper triangular matrix whose first row is
  $\bs \sigma_k^T$. Then, the contribution of user $k$ to the total distortion at subcarrier $m$ is
  \begin{align}
  P_{e,k}^{\text{(CP)}}(m) &= \EV\Bigl| \bigl[\bm Z_k^{\text{(CP)}}\bigr]_{m,n} - W_m
  H_k(\omega_m) \bigl[\bm A_k\bigr]_{m,n}\Bigr|^2 \nonumber \\
  &=2 |W_m|^2 \Re\biggl\{\bm f^H(m)(2 \bm X_k - \sigma_{k,1} \bm I_{2M})
  \Bigl[\begin{smallmatrix} \bm T_k \bm T_k^H & \bm 0\\ \bm 0 & \bm 0 \end{smallmatrix}\Bigr]
\bm f(m)\biggr\}. \label{eq:OFDM_dist}
  \end{align}
\end{proposition}
\begin{IEEEproof}
  From (\ref{eq:CP_signal_dist}), and since all the columns of $\bm A_k$ are
  i.i.d., we have
  \begin{multline*}
  P_{e,k}^{\text{(CP)}}(m) = |W_m|^2 \bm f^H(m) \widetilde{\bm H}_k \bm F_{2M}
  \EV|\bm a_k \bm a_k^H|^2 \bm F_{2M}^H \widetilde{\bm H}_k^H \bm f(m) \\{}+ 
  |W_m|^2 \bm f^H(m) \widetilde{\bm H}_{k,T} \bm F_{2M}
  \EV|\bm a_k \bm a_k^H|^2 \bm F_{2M}^H \widetilde{\bm H}_{k,T}^H \bm f(m).
  \end{multline*}
  Also, we can write
  $$
  \bm F_{2M} \EV|\bm a_k \bm a_k^H|^2 \bm F_{2M}^H = \sum_{i=1}^{2M}
  \sigma_{k,i} \biggl[\mathcal{T}_{2M}^{i-1} +
  \Bigl(\mathcal{T}_{2M}^{2M-i+1}\Bigr)^T\biggr]
  $$
  where $\mathcal{T}_{2M}$ is a $2M\times 2M$ shift matrix, with ones in the
  superdiagonal (i.e. $[\mathcal{T}_{2M}]_{i,i+1} = 1$ for all $i=1,\dots,2M-1$)
  and zeros elsewhere.

  Noting from (\ref{eq:Hk}) and (\ref{eq:HkT}) that $\widetilde{\bm H}_k$ and
  $\widetilde{\bm H}_{k,T}$ are shifted versions of the same matrix and
  recalling that $\sigma_{k,i}=\sigma_{k,2M-i+2}^*$ for all $i=2,\dots,2M$, we can write
  \begin{align*}
    \widetilde{\bm H}_k \bm F_{2M} \EV|\bm a_k \bm a_k^H|^2 \bm F_{2M}^H \widetilde{\bm H}_k^H
      &= \widetilde{\bm H}_{k,T} \bm F_{2M} \EV|\bm a_k \bm a_k^H|^2 \bm
      F_{2M}^H\widetilde{\bm H}_{k,T}^H\\
      &= \Bigl[\begin{smallmatrix} \bm T_k & \bm 0\\ \bm 0 & \bm 0 \end{smallmatrix}\Bigr]
        \bm F_{2M} \EV|\bm a_k \bm a_k^H|^2 \bm F_{2M}^H
      \Bigl[\begin{smallmatrix} \bm T_k^H & \bm 0\\ \bm 0 & \bm 0
    \end{smallmatrix}\Bigr]\\
    &= \sum_{i=1}^{2M}\sigma_{k,i}  \mathcal{T}_{2M}^{i-1}\Bigl[\begin{smallmatrix}
        \bm T_k \bm T_k^H & \bm 0\\ \bm 0 & \bm 0 \end{smallmatrix}\Bigr]
          + \sum_{i=2}^{2M}\sigma_{k,i}^*  \Bigl[\begin{smallmatrix}
        \bm T_k \bm T_k^H & \bm 0\\ \bm 0 & \bm 0
          \end{smallmatrix}\Bigr]\Bigl(\mathcal{T}_{2M}^{i-1}\Bigr)^T\\
    &= \bm X_k \Bigl[\begin{smallmatrix} \bm T_k \bm T_k^H & \bm 0\\ \bm 0 & \bm 0
          \end{smallmatrix}\Bigr] + \Bigl[\begin{smallmatrix}
          \bm T_k \bm T_k^H & \bm 0\\ \bm 0 & \bm 0 \end{smallmatrix}\Bigr] (\bm
              X_k^H - \sigma_{k,1} \Mid_{2M})
\end{align*}
since
$$
\bm X_k = \sum_{i=1}^{2M}\sigma_{k,i}  \mathcal{T}_{2M}^{i-1}.
$$
Hence, expression (\ref{eq:OFDM_dist}) follows straightforwardly.
\end{IEEEproof}

\begin{figure}
  \centering
  \begin{tabular}{r}
  \includegraphics{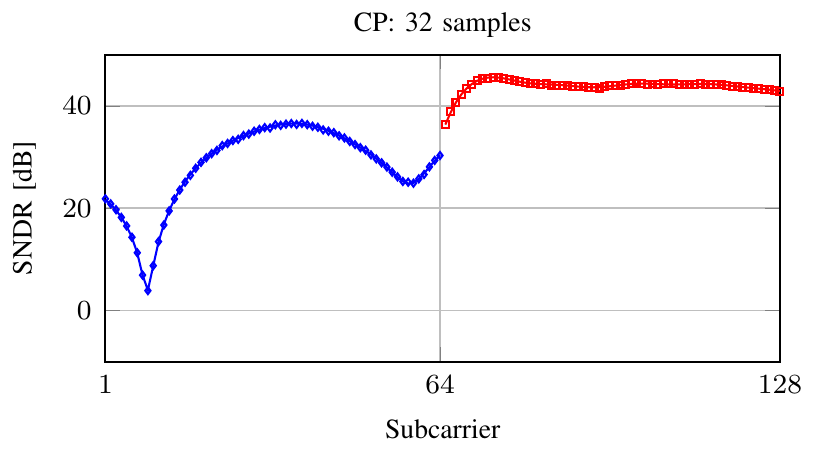}\\
  \includegraphics{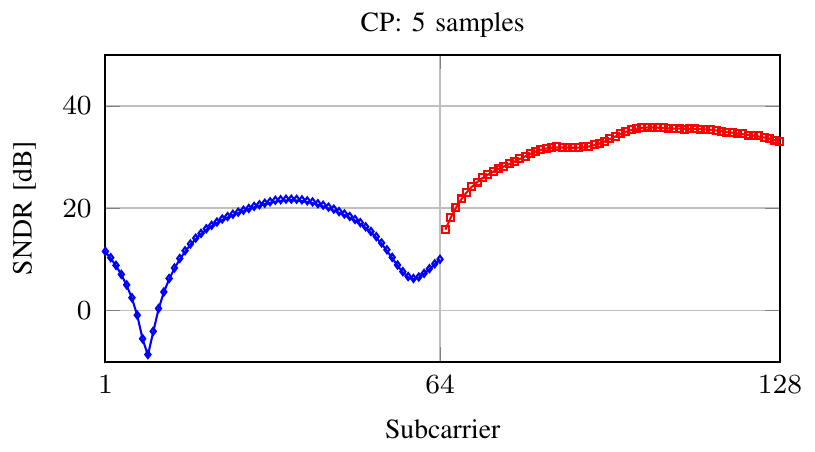}
  \end{tabular}
  \caption{Theoretic (line) and empirical (markers) SNDR for a CP-OFDMA channel with two
  users. The SNR for both users is set to 40~dB. The channel impulse responses
  are 14-tap long and correspond to the realizations of
Fig.~\ref{fig:channelsEVA}. The length of the cyclic prefix is either 32 samples
(top) or 5 samples (bottom).}\label{fig:SNDR_CP}
\end{figure}

An example of how the length of the CP affects performances is
given by Fig.~\ref{fig:SNDR_CP}, where both theoretical (line) and
empirical (markers, averaged over 2000 multicarrier symbols) values of SNDR are
reported for two different CP lengths. More specifically, for channel impulse
responses of 14 taps (corresponding to the two EVA model realizations of
Fig.~\ref{fig:channelsEVA}), we compare the SNDR obtained by the ideal case (CP
of 32 samples, top) with the one obtained by a CP of 5 samples (bottom): some
subcarriers experience losses of approximately 20 dB.

As a final remark, it is interesting to see that (\ref{eq:OFDM_dist}) takes the
form
$$
P_{e,k}^{\text{(CP)}}(m) = P_s\frac{|W_m|^2}{2M^2}\sum_{\ell=m_{\min}}^{m_{\max}}
  \frac{\sin^2\bigl[\frac{\pi}{2M}(m-\ell)(L_T-1)\bigr]}{\sin^2\bigl[\frac{\pi}{2M}(m-\ell)\bigr]}
$$
when User $k$ transmits over subcarriers $\{m_{\min},\dots,m_{\max}\}$ at power
$P_s$ and when its channel is a delay of $L_T-1$ samples (i.e., the channel
impulse response is $\delta[n-L_T]$). Note that this is the case considered in
Section~\ref{ssec:delay}.


\end{document}